\documentclass{aa}  

\usepackage{graphicx}
\usepackage{txfonts}
\usepackage{natbib}
\usepackage{gensymb}
\usepackage{tabularx}
\usepackage{subfigure}
\usepackage{longtable}
\usepackage{xtab}
\bibpunct{(}{)}{;}{a}{}{,}
\usepackage[colorlinks=true, citecolor=blue]{hyperref}
\usepackage{multirow}
\makeatletter
\renewcommand*\aa@pageof{, page \thepage{} of \pageref*{LastPage}}
\makeatother
\usepackage{etoolbox}
\makeatletter

\usepackage{multicol}

\makeatother
\usepackage{array,booktabs}

\newcommand{\alow}{$\alpha_{\text{low}}$}
\newcommand{\ahigh}{$\alpha_{\text{high}}$}
\newcommand{\Pfive}{$P_{5\,\text{GHz}}$}
\newcommand{\Vmax}{$V_{\text{max}}$}
\begin{document} 

   \title{Extragalactic Peaked-Spectrum  Radio Sources at Low-Frequencies are Young Radio Galaxies}
   \titlerunning{Peaked-Spectrum sources in LOFAR all-sky surveys}
   
   \author{M.~M.~Slob\inst{1},
        J.~R.~Callingham\inst{1,2}, H.~J.~A. R\"ottgering\inst{1}, W.~L.~Williams\inst{1}, K.~J.~Duncan\inst{3},
        F.~de~Gasperin\inst{4,5},
        M.~J.~Hardcastle\inst{6}, and G.~K.~Miley\inst{1}
          }
   \institute{Leiden Observatory, Leiden University, PO Box 9513, 2300 RA, Leiden, The Netherlands\\
        \email{slob@strw.leidenuniv.nl}
        \and
        ASTRON, Netherlands Institute for Radio Astronomy, Oude Hoogeveensedijk 4, Dwingeloo, 7991 PD, The Netherlands
        \and
        Institute for Astronomy, Royal Observatory, Blackford Hill, Edinburgh, EH9 3HJ, UK
        \and
        INAF - Istituto di Radioastronomia, via P. Gobetti 101, 40129, Bologna, Italy
        \and
        Hamburger Sternwarte, Universit\"at Hamburg, Gojenbergsweg 112, 21029, Hamburg, Germany
        \and
        Centre for Astrophysics Research, Department of Physics, Astronomy and Mathematics, University of Hertfordshire, College Lane, Hatfield AL10 9AB, UK\\
        }

   \date{Received 1 August 2022; Accepted}

\abstract{We present a sample of 373 peaked-spectrum (PS) sources with spectral peaks around 150\,MHz, selected using a subset of the two LOw Frequency ARray (LOFAR) all-sky surveys, the LOFAR Two Meter Sky Survey and the LOFAR LBA Sky Survey. These LOFAR surveys are the most sensitive low-frequency widefield surveys to date, allowing us to select low-luminosity peaked-spectrum sources. Our sample increases the number of known PS sources in our survey area by a factor 50. The 5\,GHz luminosity distribution of our PS sample shows we sample the lowest luminosity PS sources to-date by nearly an order of magnitude. Since high-frequency gigahertz-peaked spectrum sources and compact steep-spectrum sources are hypothesised to be the precursors to {large} radio galaxies, we investigate whether this is also the case for our {sample of} low-frequency PS sources. Using optical line emission criteria, we find that our PS sources are predominately high-excitation radio galaxies instead of low-excitation radio galaxies, corresponding to a quickly evolving population. We compute the radio source counts of our PS sample, and find they are scaled down by a factor of $\approx$40 compared to a {general} sample of radio-loud active galactic nuclei (AGN). This implies {that} the lifetimes of PS sources are 40 times shorter than large scale radio galaxies, if their luminosity functions are identical. To investigate this, we compute the first radio luminosity function for a homogeneously-selected PS sample. {We find that for 144\,MHz luminosities $\gtrsim 10^{25}$\,W\,Hz$^{-1}$, the PS luminosity function has the same {shape} as an unresolved radio-loud AGN population but shifted down by a factor of $\approx$10.} We interpret this as strong evidence that these high-luminosity PS sources evolve into large-scale radio-loud AGN. For local, low-luminosity PS sources, there is a surplus of PS sources, which we hypothesise to be the addition of frustrated PS sources that do not evolve into large-scale AGN.
}

\keywords{galaxies: active -- galaxies: evolution -- radio continuum: galaxies}
\authorrunning{Slob,~M.~M.~et al.}
\maketitle

%

\section{Introduction}
\label{sec:intro}
Gigahertz-peaked spectrum (GPS), compact steep spectrum (CSS), and high-frequency peaked (HFP) sources are different classes of peaked-spectrum (PS) sources. PS sources are radio-loud active galactic nuclei (AGN), defined by their small linear sizes and spectral peak in their broadband radio spectra \citep{2021A&ARv..29....3O}. The differentiation between the GPS, HFP, and CSS classes of PS sources is largely based on the frequency of their spectral peak and the maximum linear size of the source.

GPS sources are defined to have a spectral peak in the $\sim$0.4 to $\sim$5\,GHz frequency range \citep{1983A&A...123..107G, 2021A&ARv..29....3O}, while HFP sources have a spectral peak $\gtrsim$\,5\,GHz \citep{2000A&A...363..887D}. Both GPS and HFP sources have very small projected linear sizes of $\lesssim$\,1\,kpc. On the other hand, CSS sources have larger linear sizes that exceed 1\,kpc and have expected peaks below 400\,MHz \citep{1990A&A...231..333F}. There is also a newly suggested class of PS sources with observed peak-frequencies below 1\,GHz, which are sometimes referred to as megahertz peaked spectrum (MPS) sources. These MPS sources have been placed in the continuum of PS sources \citep{2004NewAR..48.1157F, 2015MNRAS.450.1477C,Callingham_2017}.

There are two hypothesised scenarios for the small linear scales and the spectral properties of peaked spectrum sources. The first of these is the \textit{youth} model, which argues that these PS sources are the precursors to massive radio-loud AGN. The \textit{youth} model is supported by the morphology of PS sources, since many display a double-lobed structure on small scales. It has been shown that a relationship between radio power and linear size of PS sources exists \citep{Kunert_Baj/j.1365-2966.2010.17271.x, 2012ApJ...760...77A}, as well as a relation between turnover frequency and linear size \citep{1997AJ....113..148O, Snellen_2000}. These relations suggest that double-lobed radio sources evolve from HFP sources into GPS sources, to CSS sources, and finally into  Fanaroff–Riley I and II (FRI and FRII) sources \citep{1985MNRAS.215..463C, Kunert_Baj/j.1365-2966.2010.17271.x}. This evolutionary model has been further supported by age estimates from spectral break modeling and observations of the motion of hot spots from high-resolution imaging{ \citep{1998A&A...337...69O,Kaiser_Best/j.1365-2966.2007.12350.x}.}

A problem with the \textit{youth} scenario is that it has been suggested that there is an overabundance of PS and CSS sources relative to large AGN \citep{1981A&AS...43..381K, 1982MNRAS.198..843P, O_Dea_1998, 2012ApJ...760...77A}. This would imply the \textit{youth} model cannot solely explain the existence of all PS sources. An alternative hypothesis is that PS sources are \textit{frustrated} -- these sources are not young galaxies but are instead confined to small spatial scales because of extremely dense gas in their central environment. The \textit{frustration} hypothesis is supported by observations of radio morphologies of CSS sources, which imply strong interactions between individual sources' radio jets and their environments \citep{1984Natur.308..619W, 1984AJ.....89....5V,Kunert_Baj/j.1365-2966.2010.17271.x}. Further evidence for the \textit{frustration} hypothesis is given by the detection of extended emission around PS sources, implying multiple epochs of activity \citep{1990A&A...232...19B, 1990A&A...233..379S}. Studies of individual sources have found evidence for unusually high densities of the surrounding medium of these sources \citep[e.g.][]{Peck_1999, Callingham_2015, Sobolewska_2019}, further supporting the \textit{frustration} hypothesis. For specific PS sources, both the \textit{youth} and the \textit{frustration} scenario may apply, since young sources with constant AGN activity could break through their dense surrounding medium \citep{2012ApJ...760...77A}. {Additionally, there is evidence that a fraction of PS sources will not grow into large AGN, but rather turn off and fade because their radio activity has stopped \citep[e.g.][]{2006A&A...450..945K,2010MNRAS.402.1892O}.}

One way to differentiate between the two imposed hypotheses is by determining the absorption mechanism that causes the spectral peak. For most PS sources a synchrotron self-absorption (SSA) model describes the observed relation between peak frequency and linear size well \citep[e.g.][]{de_Vries_2009, Snellen_2000}, while for individual sources surrounded by a dense medium a free-free absorption (FFA) mechanism fits the observed turnover better \citep[e.g.][]{Bicknell_1997, Peck_1999, Callingham_2015}. In order to differentiate between the SSA and FFA mechanisms, accurate spectral data below the turnover frequency is required. For PS sources this implies highly sensitive data at frequencies $<$\,200\,MHz is needed \citep{Snellen_2009}. Additionally, an accurate low-frequency luminosity function of PS sources would be invaluable in testing how the luminosity of PS sources evolve relative to large-scale AGN, informing us whether the youth model is compatible with the observed luminosity evolution. Low-frequency wide-field surveys provide a way to homogeneously select samples of PS sources such that we can characterise incompleteness issues that have plagued previous attempts at computing luminosity functions \citep[e.g.][]{Snellen_2000}.

Recent developments in low radio frequency telescopes have led to more reliable characterisation of low-frequency spectra and better sensitivity. Wide-field surveys from these telescopes include the GaLactic and Extragalactic All-sky Murchison Widefield Array \citep[GLEAM;][]{2015PASA...32...25W} survey, the TIFR GMRT Sky Survey \citep[TGSS;][]{2017A&A...598A..78I}, the LOFAR Two-Metre Sky Survey \citep[LoTSS;][]{2022A&A...659A...1S}, and the LOFAR LBA Sky Survey \citep[LoLLS;][]{2021A&A...648A.104D}. In particular, \citet{Callingham_2017} used the GLEAM survey to identify 1483 sources with spectral peaks between 72\,MHz to 1.4\,GHz, which doubled the number of known PS sources. Both \citet{Callingham_2017} and \citet{2019A&A...628A..56K} found that while SSA describes the turnover of a large subset of PS sources, a fraction of PS sources has to be described by a FFA model as the spectral slope below the peak violates the theoretical limit of SSA.

Recent LOw-Frequency ARray \citep[LOFAR;][]{2013A&A...556A...2V} radio surveys have started a revolution in high sensitivity data at low frequencies. More sensitive observations enable us to potentially identify high-redshift PS sources, as well as low-luminosity PS sources. Previous PS samples have median 5\,GHz radio luminosities of $10^{26} - 10^{27} \ \text{W Hz}^{-1}$ \citep[e.g.][]{O_Dea_1998, 1998A&AS..131..435S, Callingham_2017}. However, \citet{2019A&A...622A...1S} predicts that with LoTSS, PS sources with radio powers $< 10^{25} \ \text{W\,Hz}^{-1}$ can be identified. It has been argued that these low-power compact sources can be the short-lived young radio sources that could explain the overabundance of PS sources compared to radio-loud sources \citep{Kunert_Baj/j.1365-2966.2010.17271.x}. 

LOFAR has two wide-field low-frequency sky surveys ongoing. The first of these surveys, LoTSS, has had a second data release (DR2) that covers 27\% of the Northern sky at 120 -- 168 MHz, with a median sensitivity of 83\,$\mu$Jy/beam, and a resolution of 6\arcsec \citep{2022A&A...659A...1S}. The second LOFAR survey used in this study is LoLSS, which has recently had a preliminary data release (PDR) that cover 3\% of the Northern sky, with a total of 25,247 identified sources at observing frequencies between 42 -- 66 MHz, with a resolution of 47\arcsec\ and a median sensitivity of 5\,mJy/beam \citep{2021A&A...648A.104D}. The third radio survey that is important in this research is the NRAO VLA Sky Survey \citep[NVSS;][]{1998AJ....115.1693C}. NVSS covers 82\% of the Northern sky at an observational frequency of 1.4\,GHz, a resolution of 45\arcsec, and a median sensitivity of 0.45\,mJy/beam. 

The purpose of this paper is to combine observations from LoTSS, LoLSS, and NVSS to identify low-luminosity PS sources with spectral peaks at frequencies $\sim$150 MHz. We will investigate if the characteristics of these newly identified PS sources are consistent with the populations of PS sources identified by \citet{Callingham_2017}, as well as with PS source populations with spectral peaks at gigahertz-frequencies. In particular, we aim to investigate the evolution of PS sources to ascertain their role in the evolution of radio-loud AGN.

The surveys and selection criteria used to select PS sources are outlined in Sections \ref{sec:surveys} and \ref{sec:source_selec}, respectively. In Section \ref{sec:known_PS}, we cross-match our sample of PS sources to known GPS, compact-symmetric objects (CSO), and HFP sources. The 5\,GHz radio luminosities of our PS sources are presented in Section \ref{sec:radio_power}. In Section \ref{sec:HERG_LERG}, the classification of PS sources according to their black-hole accretion mechanism is outlined. Finally, we compute and analyse the radio source counts and luminosity functions in Sections \ref{sec:source_counts} and \ref{sec:LFs}, respectively. Throughout this paper we adopt the standard lambda cold dark matter cosmological model, with parameters $\Omega_{\mathrm{M}} = 0.27$, $\Omega_{\Lambda} = 0.73$, and the Hubble constant $H_0 = 70$\,km\,s$^{-1}$\,Mpc$^{-1}$ \citep{2013ApJS..208...19H}.

\section{Surveys}
\label{sec:surveys}
In order to identify PS sources, flux density measurements from at least three different radio frequencies are needed. The sensitivity and observing frequency of the surveys used to identify PS sources dictate the peak frequencies and peak flux densities of the selected PS sources. Previous PS source studies mostly were limited to using surveys above $\sim$500\,MHz, and therefore identified sources with spectral peaks in the gigahertz-frequency range \citep[e.g.][]{1998A&AS..131..435S, O_Dea_1998}. However, recently \citet{Callingham_2017} have shown that it is now possible to identify large samples of PS sources with spectral peaks around $\sim$100\,MHz .

The three surveys used to identify PS sources in this study were LoTSS, LoLSS, and NVSS. LoTSS and LoLSS are both low-frequency radio surveys conducted by LOFAR, with observing frequencies of 120--168\,MHz and 42--66\,MHz, respectively. The 1.4\,GHz NVSS survey was used as the high frequency survey. For further analysis, the PS sources were also cross-matched to other radio surveys with observational frequencies between 42 and 1400\,MHz. From the combination of surveys in this study, we are sensitive to detecting sources that have spectral peaks between $\sim$54 and $\sim$1400\,MHz.

\begin{figure}[t]
    \centering
    \resizebox{\hsize}{!}{\includegraphics{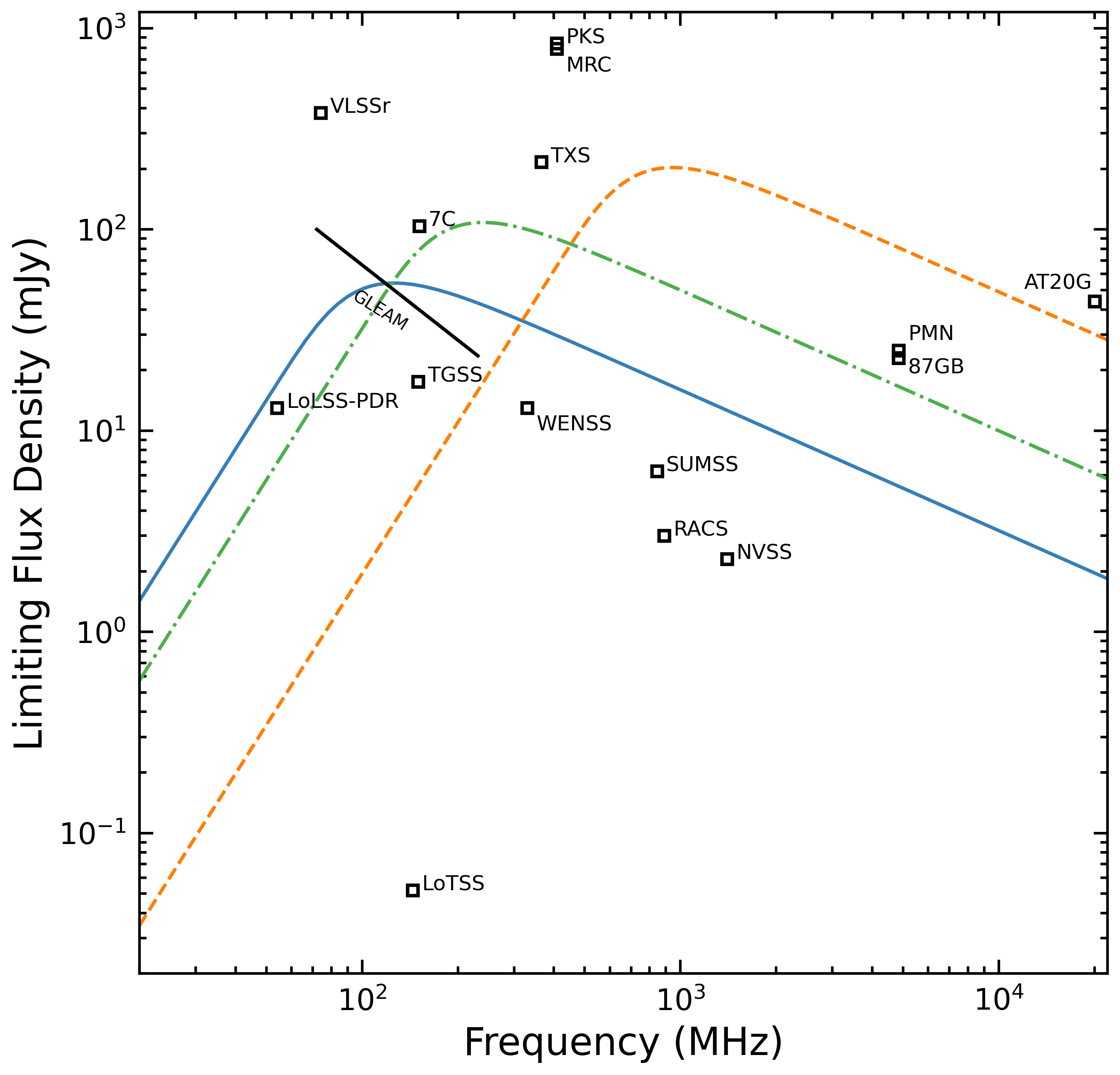}}
    \caption{The limiting flux densities and frequencies for major radio surveys. The limiting flux density for a survey is given by the faintest catalogued source flux density. The GLEAM survey is represented as a line since it has variable limiting flux densities over its range of observing frequencies. The orange dashed curve represents the observational limit for the sample presented by \citet{O_Dea_1998}. This is given by an SSA spectrum of a PS source with a peak at 750\,MHz and an observed peak flux density of 300\,mJy. The green dash-dotted curve represents the observational limit for the sample presented by \citet{Callingham_2017}. This is given by an SSA spectrum of a PS source with a peak at 190\,MHz, with a faintest observed peak flux density of 160\,mJy.
    The blue curve represents the theoretical observational limit of PS sources in this study. This is given by an SSA spectrum of a PS source with a peak at 100\,MHz, with a faintest peak flux density of 80\,mJy, based on the limiting flux densities of LoLSS-PDR, LoTSS, and NVSS. This figure also illustrates that in this study LoLSS is the survey that dictates the limiting flux density for identifying PS sources. The following plotted surveys that were not previously mentioned are: 
    Cambridge 7C \citep{2007MNRAS.382.1639H} survey, Westerbork Northern Sky Survey \citep[WENSS;][]{1997A&AS..124..259R}, Texas Survey \citep[TXS;][]{1996AJ....111.1945D}, Molonglo Reference Catalogue \citep[MRC;][]{1991Obs...111...72L}, Parkes \citep[PKS;][]{1990PKS...C......0W} survey, Sydney University Molonglo Sky Survey \citep[SUMSS;][]{2003MNRAS.342.1117M},    MIT-Green Bank 5\,GHz \citep[87GB;][]{1991ApJS...75.1011G} survey, Parkes-MIT-NRAO \citep[PMN;][]{1994ApJS...91..111W} survey, Australia Telescope 20\,GHz \citep[AT20G;][]{2010MNRAS.402.2403M} survey, and the The Rapid ASKAP Continuum Survey \citep[RACS;][]{69292890e76147ccbcc8652c97fc8d03}.}
    \label{fig:survey_limits}
\end{figure}

Of the three surveys used to identify PS sources in this research, LoLSS has the lowest sensitivity. Since LoLSS will provide the low-frequency data-point for identifying PS sources, the completeness limit of our PS source sample will largely be set by the sensitivity of LoLSS. Besides this, LoLSS PDR does not currently cover the full observation area of LoTSS DR2 and NVSS, which means not all of the available coverage of LoTSS DR2 and NVSS will be used in this study. We show in Fig. \ref{fig:survey_limits} the PS spectra that this study and previous studies identify, based on the limiting flux densities of radio surveys.

\subsection{LOFAR surveys}
\label{sec:LoTSS_LoLSS}
Two new LOFAR surveys formed the basis of our study -- LoTSS and LoLSS. LOFAR is a radio interferometer with 52 dipole-antenna stations across the Netherlands and Europe. Each LOFAR station consists of low- and high-band antennae (LBA and HBA, respectively), which are used for 10--90\,MHz and 110--250\,MHz observations, respectively \citep{2013A&A...556A...2V}.

The first LOFAR survey that was used in this study is LoTSS-DR2 \citep{2022A&A...659A...1S}. LoTSS-DR2 was formed from observations taken by LOFAR at 120 -- 168\,MHz between 2014-05-23 and 2020-02-05. This survey consists of 4,395,448 catalogued sources from two regions centered at 12h45m00s +44\textdegree30\arcmin00\arcsec\ and 1h00m00s +28\textdegree00\arcmin00\arcsec\ spanning 4178 and 1457 square degrees, respectively. In total this survey covers 27\% of the Northern sky. LoTSS has a 6\arcsec\ resolution, and is 90\% complete at 0.8 mJy \citep{2022A&A...659A...1S}. The full data reduction process for LoTSS is described in detail by \citet{2022A&A...659A...1S}. For the sources detected in LoTSS, there is an optical catalogue available comprised of two non-overlapping optical catalogues from LoTSS Data Release 1 \citep[DR1;][]{2019A&A...622A...2W, 2019A&A...622A...3D} and LoTSS Data Release 2 \citep[DR2, Hardcastle et al., in prep.,][]{2022MNRAS.512.3662D}. This ancillary optical database provides us with optical source associations and multi-wavelength properties of the identified PS sources, including photometric and spectroscopic redshifts.

For sources in LoTSS-DR2 in-band spectra were also created. These in-band spectra consist of three flux density measurements with 16\,MHz bandwidth and central frequencies of 128, 144 and 160\,MHz. However, as shown by \citet{2022A&A...659A...1S}, these in-band spectra are not reliable for most sources. We have therefore not used these in-band spectra for spectral modeling or identifying PS sources in this study. However, the in-band spectra are plotted in the spectral energy distributions to act as a visual guide in determining the {reliability} of a spectral fit.

The second LOFAR survey used in this research is the preliminary data release of LoLSS \citep{2021A&A...648A.104D}, with observations at 42 -- 66\,MHz, formed from observations conducted between 2017 and 2019. This survey consists of 25,247 sources centered around the Hobby-Eberly Telescope Dark Energy Experiment (HETDEX) Spring Field with Right Ascensions from 11 to 16\,h and Declinations from 45\textdegree\ to 62\textdegree \citep{2008ASPC..399..115H}, covering 740 square degrees. LoLSS has an angular resolution of 47\arcsec, and is 90\% complete at 40 mJy. Since LoLSS is the survey with the smallest survey area, its sky footprint corresponds to the detection area of this study. In the final data release of LoLSS, a higher resolution of 15\arcsec with better sensitivities of 1--2\,mJy will be reached (de Gasperin et al., in prep.).

\subsection{NVSS}
\label{sec:NVSS}
NVSS is the high frequency radio survey we used to identify PS sources in this study. NVSS is a continuum survey formed from observations conducted by the Very Large Array (VLA) at 1.4\,GHz between September 1993 and October 1996 \citep{1998AJ....115.1693C}. The NVSS catalogue consists of 1,773,484 sources north of a declination of -40\textdegree, covering 82\% of the Northern sky. NVSS has an angular resolution of 45\arcsec\ and is 99\% complete at 3.4\,mJy.

\subsection{Additional radio surveys}
\label{sec:other_radio}
In this study we used LoTSS, LoLSS, and NVSS to identify PS sources. However, after the initial identification of these PS sources, we also cross-matched these sources to other wide-field radio surveys. The additional radio surveys used in this research are the Very Large Array Low-Frequency Sky Survey Redux \citep[VLSSr;][]{2014MNRAS.440..327L} at 74\,MHz, the TIFR GMRT Sky Survey Alternative Data Release 1 \citep[TGSS-ADR1;][]{2017A&A...598A..78I} at 150\,MHz, and the Faint Images of the Radio Sky at Twenty-cm \citep[FIRST;][]{1995ApJ...450..559B} survey at 1.4\,GHz. If a PS source had counterparts in VLSSr and/or TGSS, the flux densities of these surveys were used for spectral modeling of the source. The source counterparts from the FIRST survey were only used as visual guides in the spectral energy distribution plots to confirm the accuracy of spectral fits to the other surveys.

\section{PS source selection}
\label{sec:source_selec}
Our PS source sample has been formed by making cuts based on resolution, isolation, and whether a peak occurs in their spectra. The different selection criteria, and the number of sources left in our sample after each cut, are summarized in Table \ref{tab:source_selec}. Details and justifications of each selection step are provided below.

\begin{table}[t]
    \small
    \centering
    \caption{Summary of the selection criteria used, and the number of sources left after each selection step.}
    \begin{tabularx}{\hsize}{p{0.15\hsize} X p{0.15\hsize}}
    \hline\hline
    Selection Step & Selection Criterion & Number of Sources\\
    \hline
    0 & Total LoTSS catalogue & 4,396,228 \\
    1 & Isolated from other sources in 47\arcsec\ radius in LoTSS & 2,676,735\\
    2 & Unresolved in LoTSS & 2,493,574 \\
    3 & Sources classified as `Simple' or `Multiple' by \texttt{PyBDSF} & 2,493,565\\
    4 & Master Sample \newline LoLSS and NVSS counterparts & 9,768 \\
    5 & PS sample \newline \alow\,$\geq$ 0.1 and \ahigh\,$\leq 0$  & 373\\
    5a & Hard sample \newline \alow\,$\geq$ 0.1 and \ahigh\,$\leq -0.5$ & \textit{212} \\
    5b & Soft sample \newline \alow\,$\geq$ 0.1 and $0 \geq$\,\ahigh\,$\geq -0.5$ & \textit{161} \\ 
    \hline\hline
    \end{tabularx}
    \tablefoot{The details of each selection step are provided in Section \ref{sec:source_selec}. The italicised numbers in step 5a and 5b indicate the subset of sources selected from the sample of sources in step 5. {The large decrease of sources from selection step 3 to selection step 4 is caused in part by the fact that the LoLSS survey area only contains $\sim$20\% of the total number of sources from the LoTSS catalogue.}}
    \label{tab:source_selec}
\end{table}

\subsection{Source isolation, resolution, and cross-matching}
\begin{enumerate}
    \item To ensure the derived radio spectra were not impacted by source confusion, any source that was not isolated was removed from our sample. A LoTSS source was deemed isolated if it had no other source within a 47\arcsec\ radius. This isolation radius corresponds to the angular resolution of LoLSS, the lowest of all the surveys used in this study. A 47\arcsec\ isolation radius ensures that sources in LoLSS are not composed of multiple, independent LoTSS sources. 
    
    However, some bright ($\gtrsim$\,200 mJy) sources have deconvolution errors introduced by the data reduction pipeline of LoTSS, which give rise to nearby, incorrectly catalogued sources. To ensure we did not remove bright sources from the final sample due to deconvolution errors, the flux densities of nearby sources were also considered before flagging a source as not being isolated. If a neighbouring source had a flux density $\leq$\,10\% of the central source, the source was deemed as isolated. Even if the faint source is a not a deconvolution artefact, a flux density $\lesssim$10\% relative to the bright source means the impact on the final spectrum is negligible. This isolation selection criteria reduced the sample to roughly 60\% of the total LoTSS catalogue, to 2,676,735 sources.
    
    \item Since PS sources have small spatial scales ($\lesssim$\,1$''$, e.g. \citet{O_Dea_1998, 2018MNRAS.474.4937C}), all sources that are resolved in LoTSS were removed from our sample as it implies the source has structure $>$\,6$''$. \citet{2022A&A...659A...1S} have defined a criterion for identifying resolved sources in LoTSS. This criterion identifies a source as resolved when the natural logarithm of the ratio of the integrated flux density ($S_{\text{I}}$) to peak flux density ($S_{\text{P}}$), given by $R = \ln S_{\text{I}} /S_{\text{P}}$, is greater than or equal to $R_{99.9}$, given by
    
    \begin{equation}
        \label{eq:R99}
        R_{99.9} = 0.42 + \left( \frac{1.08}{1+\left(\frac{SNR}{96.57} \right)^{2.49}} \right)\,,
    \end{equation}
    
    where SNR is the Signal-to-Noise Ratio, defined as $\frac{S_{\text{I}}}{\sigma_{\text{I}}}$, with $\sigma_{\text{I}}$ the statistical error on the integrated flux density. All sources with $R \geq R_{99.9}$ were classified as resolved, and removed from our sample. This removed $\approx$6.8\% sources from the previous cut, leaving us with 2,493,574 sources.
    
    \item The above unresolved cut has removed most resolved sources. However, since the resolved criterion is based on the 99.9 percentile of a distribution ($R_{99.9}$), and the isolated criterion has a flux density cut, we need to include another selection step to ensure all unresolved, non-isolated sources are removed. This was done by removing all sources classified in the LoTSS catalogue as `C' type by \texttt{PyBDSF} \citep{2015ascl.soft02007M}. These are sources fit by a single Gaussian, but within an island of emission that contains other sources - such as radio relics around clusters. The remaining sample only contains sources classified by \texttt{PyBDSF} as `S' and `M' type sources. The `S' type sources are isolated sources fitted with a single Gaussian, while the `M' type sources are fitted with multiple Gaussians. These `M' type sources have not been removed from our sample since the extended emission of these sources is mostly caused by deconvolution errors around bright sources. The multiple Gaussians fitted to these sources thus generally do not represent actual extended emission. A total of 9 `C' type sources were removed from our sample.
    
    \item The remaining sources from LoTSS were then cross-matched to LoLSS, NVSS, VLSSr, TGSS and FIRST. This was done using the Tool for OPerations on Catalogues And Tables (TOPCAT, \citet{2005ASPC..347...29T}) Starlink Tables Infrastructure Library Tool Set (STILTS) multi-table cross-matching tool. A cross-matching radius of 15\arcsec\ was used for all surveys. As described in Section \ref{sec:surveys}, cross-matches in LoLSS, LoTSS, and NVSS are required to identify whether a source is a PS source. We therefore remove all sources that do not have a cross-match in LoLSS or NVSS from our sample. Since LoLSS has a smaller sky coverage than LoTSS, and both LoLSS and NVSS have lower sensitivity than LoTSS, only $\approx$0.4\% of the isolated, unresolved LoTSS sources have counterparts in both surveys. After this crossmatching step we are left with 9,768 sources in our sample, from which we can identify PS sources. We will refer to this sample of sources as the Master sample.
\end{enumerate}

\subsection{Spectral classification}
From the Master sample, PS sources were identified by their spectra. A defining feature of PS sources are the power-law components above and below the spectral peak. Firstly, we fit a generic non-thermal power-law model of the form

\begin{equation}
\label{eq:simple_pow_law}
    S_{\nu} = a\nu^{\alpha}\,,
\end{equation} 

\noindent where $S_{\nu}$ is flux density, $a$ is the amplitude of the synchrotron emission in Jy, $\nu$ is the frequency in MHz, and $\alpha$ is the spectral index. This model was fit to the flux densities from LoLSS and LoTSS to obtain the low-frequency spectral index \alow. The power-law model was also fit to the flux densities of LoTSS and NVSS to obtain the high frequency spectral index \ahigh. Since our fitting routine uses too few data-points to calculate the uncertainties on \alow\ and \ahigh, we estimate these uncertainties by fitting a power-law to the 1$\sigma$ upper- and lower uncertainties of the flux densities from one survey to the lower- and upper uncertainties of the second survey, respectively. The corresponding limiting values of \alow\ and \ahigh\ for these fits then represent the 1$\sigma$ limits of each respective \alow\ and \ahigh.

\begin{figure*}[t]
    \centering
    \includegraphics[width=17cm]{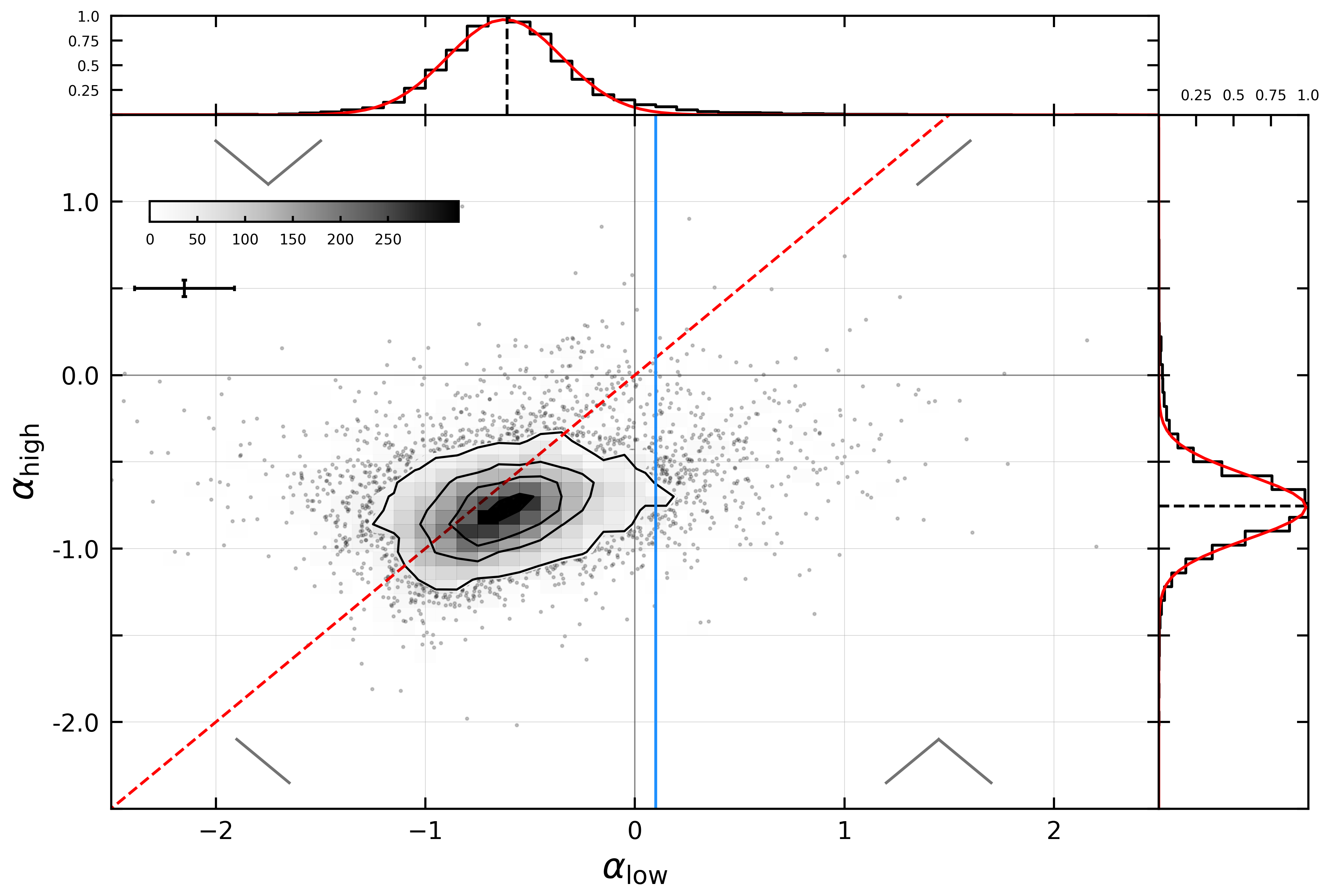}
    \caption{Radio colour-colour diagram for the 9,768 LoTSS sources left in our Master sample. \alow\ represents the spectral index between the LoLSS (54\,MHz) and LoTSS (144\,MHz) flux density points. \ahigh\ represents the spectral index between the LoTSS and NVSS flux density points. To better illustrate the large number of sources around the median of \alow\ and \ahigh, given by (\alow, \ahigh) $= (-0.6 \pm 0.2, -0.8 \pm 0.1)$, contours and a density map are plotted in this region. The contour levels represent 15, 75, 145, and 265 sources, respectively. The colours in the underlying density map show the number of sources for each pixel. The number of sources corresponding to each corresponding shading are illustrated by the colour bar at the top left of the plot. The grey lines show spectral indices of zero, to illustrate the four quadrants of the plot. The dashed red line represents a one-to-one relation between \alow\ and \ahigh. The blue line at \alow = 0.1 illustrates the selection limit used to identify PS sources in this study. In the corner of each quadrant the shape of a typical spectrum in that quadrant is shown in grey. To avoid confusion, individual error bars are not plotted, but in the top left of the plot the median error bar size is shown. The histograms at the top and right of the diagram illustrate the distributions of \alow\ and \ahigh, respectively. These distributions have been normalised to the maximum value in the distribution. In these histograms, Gaussian fits to these distributions are overplotted in red, and the dashed black lines show the median values for these distributions.}
    \label{fig:col-col}
\end{figure*}

All sources can now be situated in radio colour-colour phase space, as given by the \alow\ and \ahigh\ of the sources. In this radio colour-colour phase space, spectral peaks can be identified \citep[e.g.][]{2006MNRAS.371..898S, Callingham_2017}. The radio colour-colour phase space for the 9,768 sources in our Master sample, as obtained from the flux density points of LoLSS, LoTSS and NVSS, is presented in Fig. \ref{fig:col-col}. As expected, most sources cluster around a median of (\alow, \ahigh) $= (-0.6 \pm 0.2, -0.8 \pm 0.1)$ in the third quadrant of Fig. \ref{fig:col-col}. This is consistent with previous spectral index studies at similar observing frequencies, although our sample has a higher median value and a larger semi-interquartile range for \alow\ than previous studies \citep[e.g.][]{2007A&A...471.1105T, 2014MNRAS.440..327L, Callingham_2017}. This larger standard deviation in \alow\ is likely due to larger uncertainties in calibrating the flux density scales for LoLSS compared to the higher-frequency surveys that were used in previous studies \citep{2021A&A...648A.104D}.

Sources in the third quadrant of Fig. \ref{fig:col-col} have spectra that are described by an optically thin synchrotron power law. Sources in the first quadrant of Fig. \ref{fig:col-col} follow a positive power law from 54 MHz to 1.4\,GHz. Such sources are expected to have a spectral peak in the gigahertz-range consistent with archetypal GPS sources. Sources in the fourth quadrant have convex spectra. These sources could have another spectral turnover at a frequency $\gtrsim$\,1\,GHz, which could indicate multiple epochs of AGN activity. Sources with a spectral turnover between 54\,MHz and 1400\,MHz are located in the second quadrant of Fig. \ref{fig:col-col}. The PS sources we are interested in are therefore selected from this section of the radio colour-colour phase space.

\begin{figure}[t]
    \centering
    \resizebox{\hsize}{!}{\includegraphics{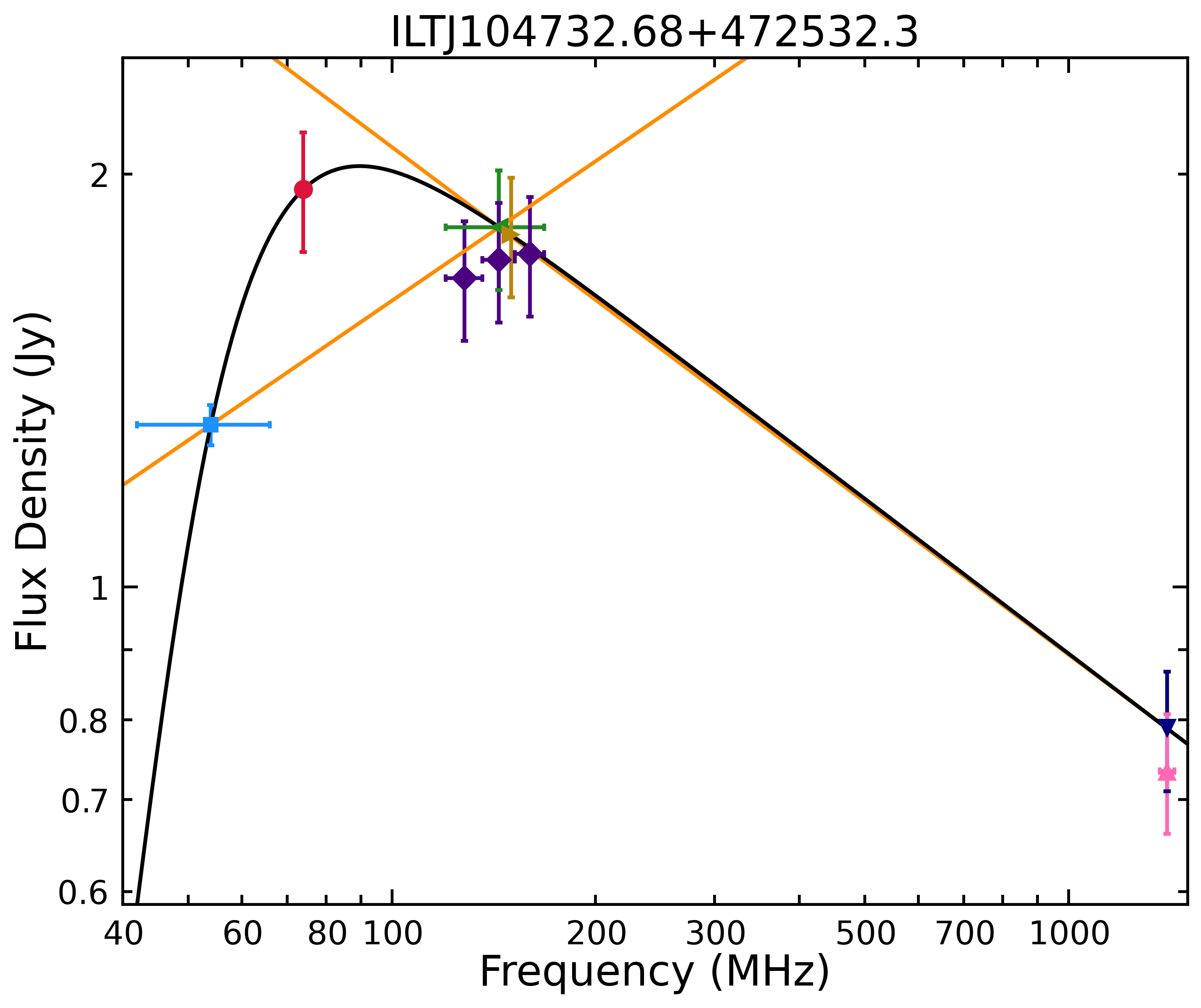}}
    \resizebox{\hsize}{!}{\includegraphics{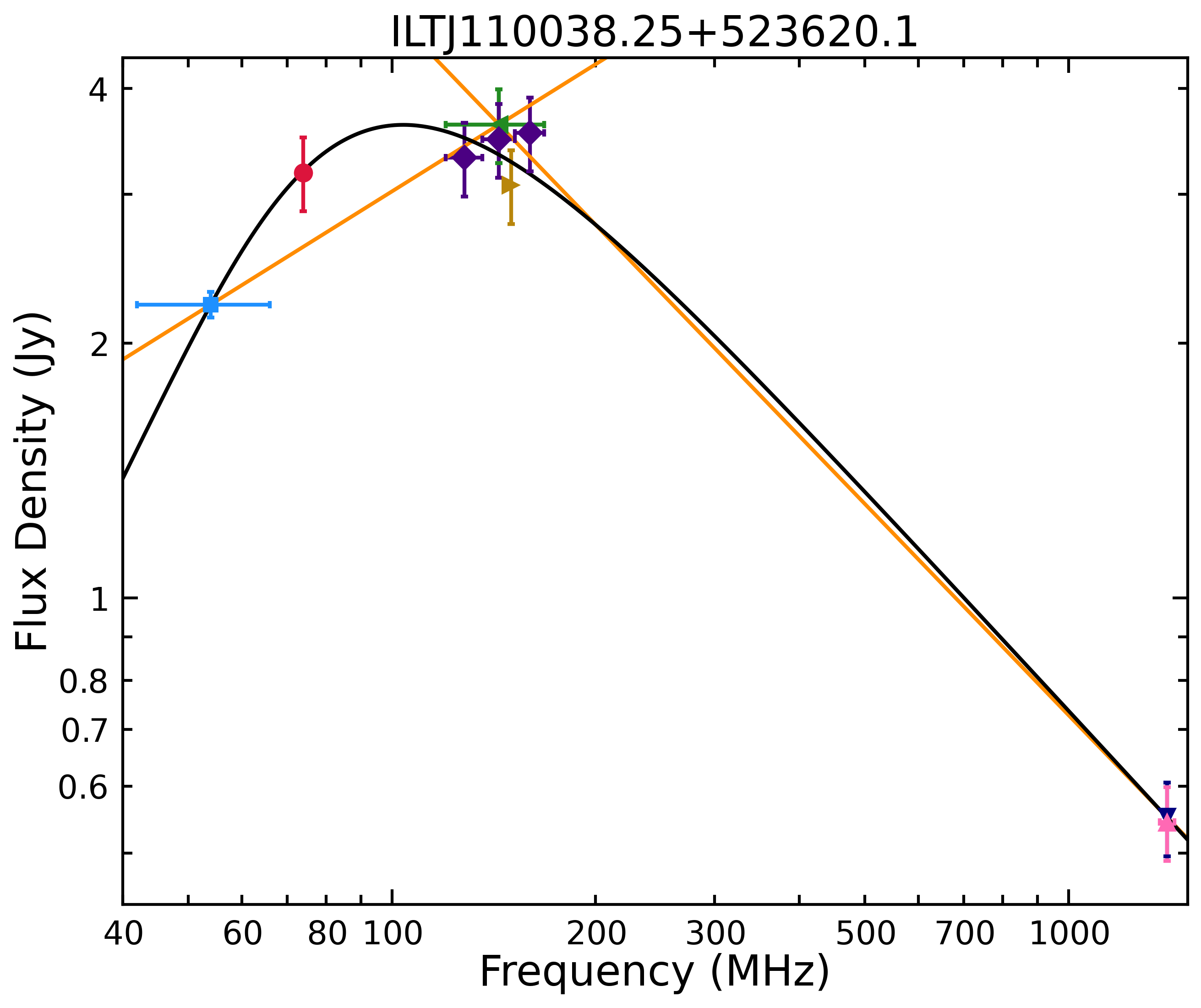}}
    \caption{Example spectra of a soft sample PS source (top) and a hard sample PS source (bottom). The flux densities of LoLSS (blue square), VLSSr (red circle), LoTSS-inband spectra (purple diamonds), LoTSS (green leftward-pointing triangle), TGSS (brown rightward-pointing triangle), NVSS (indigo downward-pointing triangle), and FIRST (pink upward-triangle) are plotted where available. The black curve represents the generic curved model from Equation \eqref{eq:curve_mod}, which was fitted to the LoLSS, LoTSS, VLSSr, TGSS and NVSS datapoints. The power-law fits used to determine \alow\ and \ahigh\ are shown in orange.}
    \label{fig:soft_hard_sed}
\end{figure}

\begin{enumerate}
\setcounter{enumi}{4}
    \item Sources in the second quadrant have a peak in their spectrum around 144\,MHz. Not all sources in this second quadrant have been classified as PS sources, but instead the cut for a source to be a PS source was made at \alow\,$\geq$ 0.1. This cut minimises the contamination of flat spectrum sources in the selected PS sample. In the literature, most previous studies have also made a cut for PS sources at \ahigh\,$\leq -0.5$ \citep{O_Dea_1998}. However, from the continuous distribution of \ahigh\ around \ahigh\,$=-0.5$ in Fig. \ref{fig:col-col}, this limit appears to be arbitrary, as also concluded by \citet{Callingham_2017}. In order to compare the results from this study to previous studies, we have made a distinction between a hard PS sample, containing PS sources with \ahigh\,$\leq -0.5$, and a soft PS sample, containing PS sources with $0.0 \geq$\,\ahigh\,$\geq -0.5$. The hard PS sample contains a total of 212 sources, and the soft PS sample contains 161 sources. The full PS sample is obtained from the combination of the soft and hard PS samples, and contains 373 sources. The spectra of a source from the soft and hard samples are shown in Fig. \ref{fig:soft_hard_sed}. The table with the optical and radio characteristics for the PS sample is available online in the style of the table presented in Appendix \ref{ap:PS_table}.
\end{enumerate}

From applying these selection steps, we have identified a sample of 373 candidate PS sources. To verify the peak in the spectra of these sources, and to better sample their spectra, we cross-matched these sources to VLSSr and TGSS. With these additional spectral points, a curved model was fit to the spectra of a sub-sample of our PS sources using the least-squares method. The generic curved model used for this is of the form
\begin{equation}
    \label{eq:curve_mod}
    S_{\nu} = \frac{S_{\text{p}}}{\left(1- e^{-1} \right)} \left(1 - e^{-\left(\nu / \nu_p\right)^{\,\alpha_{\text{thin}} - \,\alpha_{\text{thick}}}}  \right) \left( \frac{\nu}{\nu_{\text{p}}} \right)^{\alpha_{\text{thick}}} ,
\end{equation}
where $S_{\nu}$ is the flux density at frequency $\nu$, in MHz. $S_{\text{p}}$ is the flux density at the peak frequency $\nu_{\text{p}}$. $\alpha_{\text{thin}}$ and $\alpha_{\text{thick}}$ are the spectral indices in the optically thin and optically thick parts of the spectrum, respectively \citep{1998A&AS..131..435S}.

Since this model depends on four parameters, a cross-match to at least two more surveys besides LoLSS, LoTSS, and NVSS was needed to obtain a fit for this model. In total, 36 out of 373 PS sources had enough cross-matches to accurately fit a curved model. Note we do not use the results of the curved spectral model for PS sources for further statistical analysis in this paper but the curved spectral model can provide an extra confirmation for the identification of the PS nature for an individual source.

\subsection{Flux density completeness of PS sample}
\label{sec:PS_completeness}
For further analysis of the population, the flux density completeness of the PS sample needs to be known. Since our radio detection limit is dominated by the LoLSS flux density limit, we use this limit to compute the flux density limit at 144\,MHz for our sample. We extrapolate the LoLSS 90\% completeness limit at 54\,MHz of $S_{\mathrm{54\,MHz, 90\%}}= 40$\,mJy \citep{2021A&A...648A.104D} to 144\,MHz using a simple power law to find the flux limit at this selection frequency. For the PS sample, we use a spectral index for extrapolation of $\alpha_{\text{low}} = 0.1$, corresponding to the selection limit of PS sources. This results in an estimated limiting flux density for our PS sample of $S_{\text{144\,MHz, PS lim}} = 44$\,mJy. This means a LoTSS source can only be identified as peaked-spectrum in our analysis if it has a LoTSS flux density exceeding 44\,mJy.

\subsection{Redshift information}
\label{sec:redshift_info}
\begin{figure}[t]
    \centering
     \resizebox{\hsize}{!}{\includegraphics{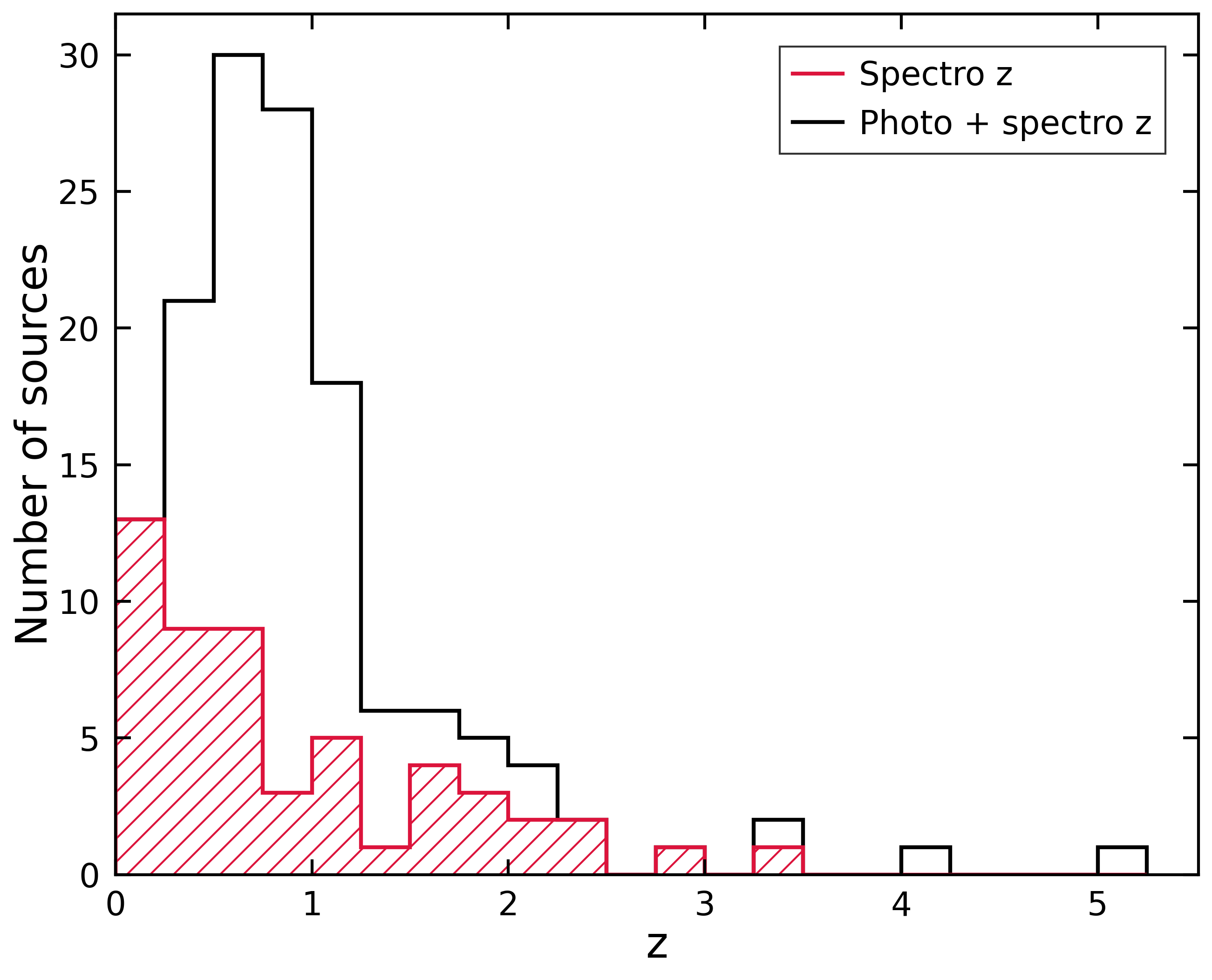}}
    \caption{Redshift distribution of the 138 PS sources with available redshift data. The red distribution illustrates the distribution of the spectroscopic redshifts which were available for 54 sources. The black distribution illustrates the distribution of the best redshifts, from the combination of spectroscopic and photometric redshifts.}
    \label{fig:z_dist}
\end{figure}

For the Master sample, redshift information from the LoTSS optical catalogues was obtained. There were two separate, non-overlapping optical catalogues available for Data Release 1 \citep[DR1; 2019][]{2019A&A...622A...2W} and DR2 (Hardcastle et al., in prep.) of LoTSS. The photometric redshifts in these optical catalogues were obtained using the methods outlined by \citet{2019A&A...622A...3D} for DR1 and \citet{2022MNRAS.512.3662D} for DR2. For sources in DR1, the spectroscopic redshift and the median photometric redshift were used. For sources in DR2, the spectroscopic redshift from Sloan Digital Sky Survey (SDSS) and the estimated photometric redshift were used if they were flagged as good-quality. For both DR1 and DR2, the spectroscopic redshift was preferred when available. In total, for 5,026 sources in our Master sample optical counterparts were identified. This resulted in available redshift data for 3,303 out of 9,768 sources in the Master sample. For the sources classified as PS sources in selection step 5, redshift information is available for 138 out of 373 sources. Spectroscopic redshift information was available for 54 sources in our PS sample. The corresponding redshift distribution for the PS sources is shown in Fig. \ref{fig:z_dist}. This distribution has a median redshift of 0.80, and a highest redshift of 5.01.

\section{Cross-matching to known PS samples}
\label{sec:known_PS}
We test the reliability of our PS source selection criteria by checking if our sample contains previously identified GPS, CSO, and HFP sources. This sample of previously known GPS, CSO, and HFP sources was obtained by collating the known GPS, CSS, and HFP source samples described by \citet{Callingham_2017}, and removing the CSS sources from this sample. Our final known GPS, CSO, and HFP sample consists of the samples isolated by \citet{O_Dea_1998}, \citet{1998A&AS..131..435S}, \citet{2000ApJ...534...90P}, \citet{2002MNRAS.337..981S}, \citet{2005A&A...432...31T}, \citet{2007A&A...463...97L}, \citet{2004A&A...424...91E}, and \citet{2011MNRAS.416.1135R}. In this sample of known PS sources, seven sources are located in the survey area of LoLSS used in this study. Of these seven sources, two have counterparts in our Master sample. The remaining five sources are too faint to be detected in LoLSS. Of these two known PS sources, neither where identified via our PS sample criteria. 

\begin{figure*}[t]
    \centering
     \subfigure{\includegraphics[width = 9.15cm]{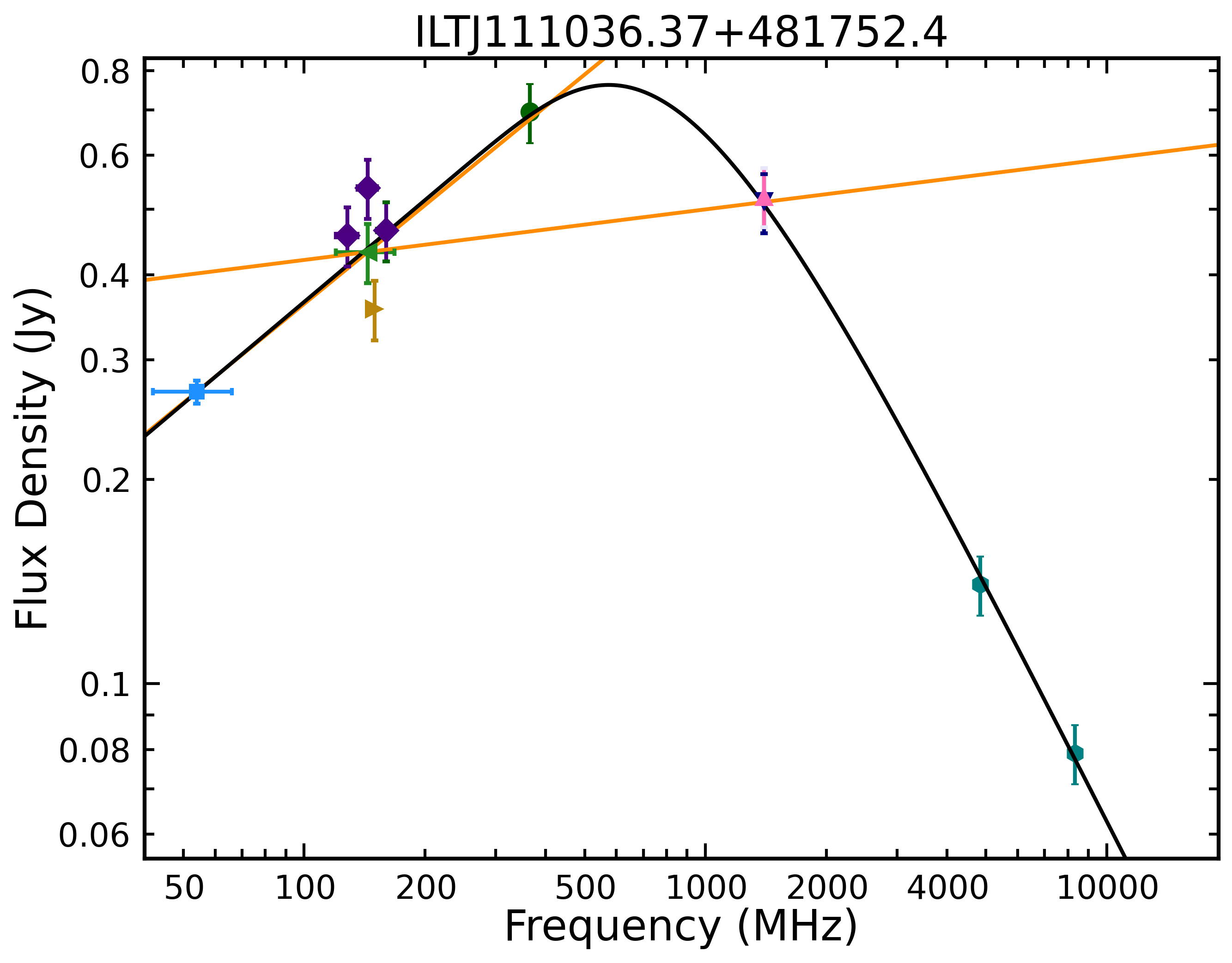}}
     \subfigure{\includegraphics[width = 9cm]{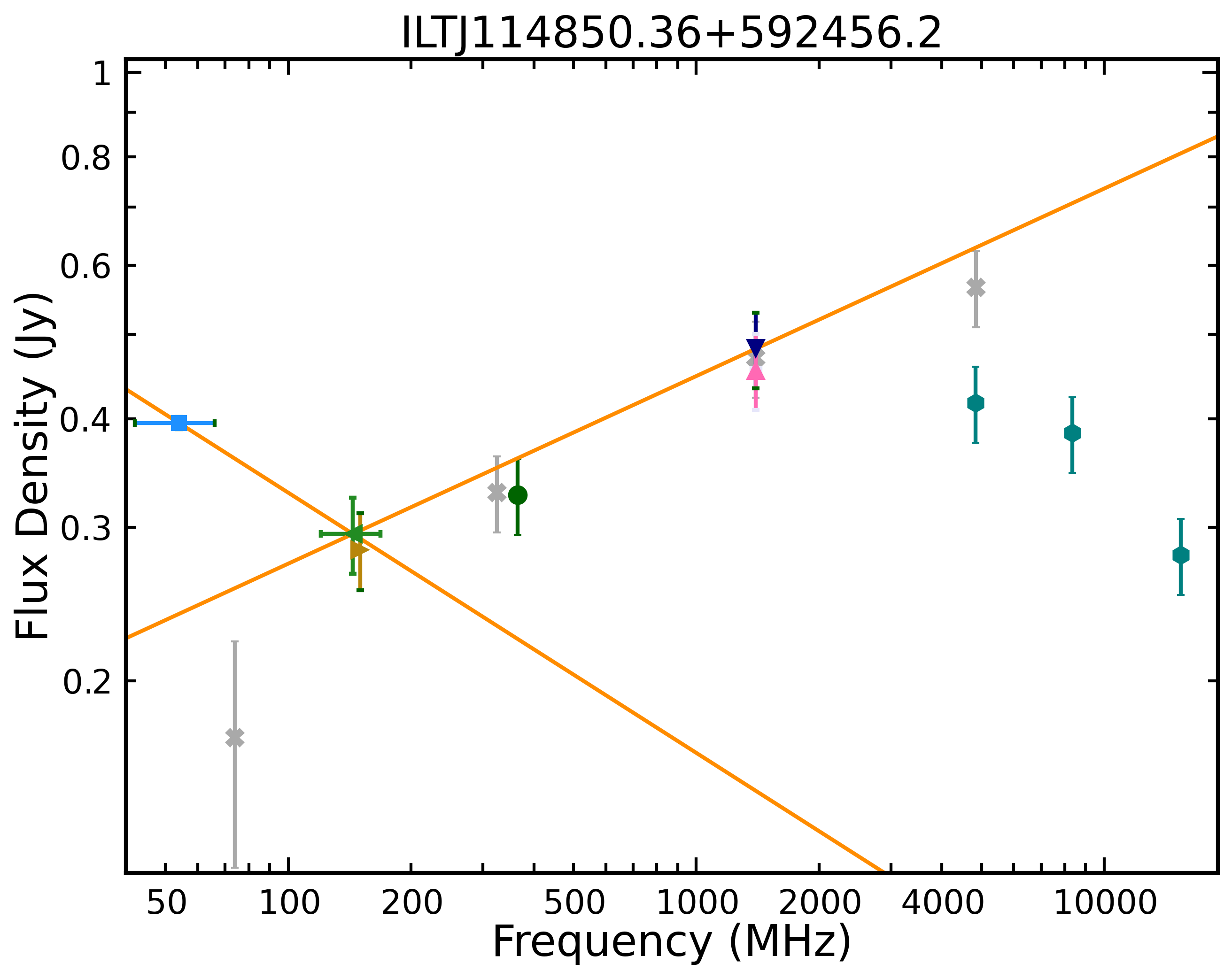}}
    \caption{The spectra of the two known GPS, CSO, and HFP sources in the Master Sample. The symbols represent data from the same surveys as in Fig. \ref{fig:soft_hard_sed}, with the grey crosses showing data presented by \citet{2015AJ....149...32M}, the darkgreen circles showing data from the TXS survey \citep{1996AJ....111.1945D}, and the teal hexagons depicting data presented by \citet{2016MNRAS.459..820T}. The black curve in the left plot shows the generic curved model from Equation \eqref{eq:curve_mod}.} 
    \label{fig:cross_PS_1}
\end{figure*}

To understand why neither of these two sources were selected by our PS sample selection, we inspect the spectral properties by characterising their radio spectra between 54\,MHz and 20\,GHz. The spectra of these two sources are shown in Fig. \ref{fig:cross_PS_1}. 

The source ILT\,J111036.37+481752.4 has a counterpart in the CSO sample presented by \citet{2000ApJ...534...90P}, and has a peaked spectrum near $\approx$250\,MHz. However, using our spectral fitting routine described in Section \ref{sec:source_selec}, we computed the spectral indices of this source to be \alow = 0.48 $\pm$ 0.14 and \ahigh = 0.07 $\pm$ 0.05, thus not classifying it as a PS source. This is because the spectrum appears flat between 144\,MHz and 1.4\,GHz. The source has an abnormally large spectral width of $\approx$1.8\,GHz - which is substantially larger than the median FWHM of 750\,MHz of the PS sample isolated by \citet{Callingham_2017}. 

The source ILT\,J114850.36+592456.2 also has a counterpart in the CSO sample presented by \citet{2000ApJ...534...90P}. From Fig. \ref{fig:cross_PS_1}, we see that this source potentially has a convex spectrum between 54--1400\,MHz, with a possible spectral turnover at $\sim$4\,GHz. However, since spectral variability can be significant at high frequencies, more high-frequency data-points after the turnover are needed in order to confirm this spectral turnover. A convex source with a spectral turnover above 1\,GHz is suggestive of multiple epochs of AGN activity, where the low-frequency section of the spectrum is dominated by emission from aged electrons, while the higher-frequency peak comes from recent core activity \citep[][and references therein]{Callingham_2017}.

In summary, for the two known PS sources that were in our Master sample, one had a convex spectrum at our selection frequencies and the other had a relatively flat peak between our LoTSS and NVSS data-points. The latter demonstrates that our PS sample will thus be a slight underestimation of the total number of PS sources in the Master sample, especially to those with wide spectral peaks. We can also conclude that the 373 sources in our PS sample are all newly identified PS candidates, increasing the number of known PS sources in our detection area by a factor of 50.

\section{5 GHz radio powers}
\label{sec:radio_power}
To investigate how the radio luminosity distribution of our PS sample compares to literature PS samples, we computed the 5\,GHz luminosity of the 3,303 sources in the Master sample that have redshift information available. Out of these 3,303 sources, 138 sources are PS sources. The frequency of 5\,GHz was chosen in order to compare the luminosities of our sample with the samples of \citet{O_Dea_1998}, \citet{1998A&AS..131..435S}, and \citet{Callingham_2017}, who all evaluated their radio luminosities at 5\,GHz. The 5\,GHz radio luminosity, \Pfive, was computed using
\begin{equation}
    \label{eq:P5GHz}
    \text{P}_{\text{5\,GHz}} = \frac{4\pi D^2_L S_{\text{5\,GHz}}}{(1+z)^{1+\text{\ahigh}}},
\end{equation}
where $D_L$ is the luminosity distance, $S_{\text{5\,GHz}}$ is the flux density of a source at 5\,GHz, and the factor $1/(1+z)^{1+\text{\ahigh}}$ is the $k$-correction. We computed $S_{\text{5\,GHz}}$ by extrapolating the power-law from equation \eqref{eq:simple_pow_law} up to 5\,GHz with spectral index \ahigh. This assumes the spectrum of a source follows the same power-law fitted between 140\,MHz and 1.4\,GHz, up to frequencies of 5\,GHz, without strong deviations. We expect this assumption to be valid, since deviations due to spectral curvature are not significant for our sources at frequencies above 1\,GHz but below 10\,GHz \citep{2012MNRAS.422.2274C,Callingham_2017}.

\begin{figure}[t]
    \centering
     \resizebox{\hsize}{!}{\includegraphics{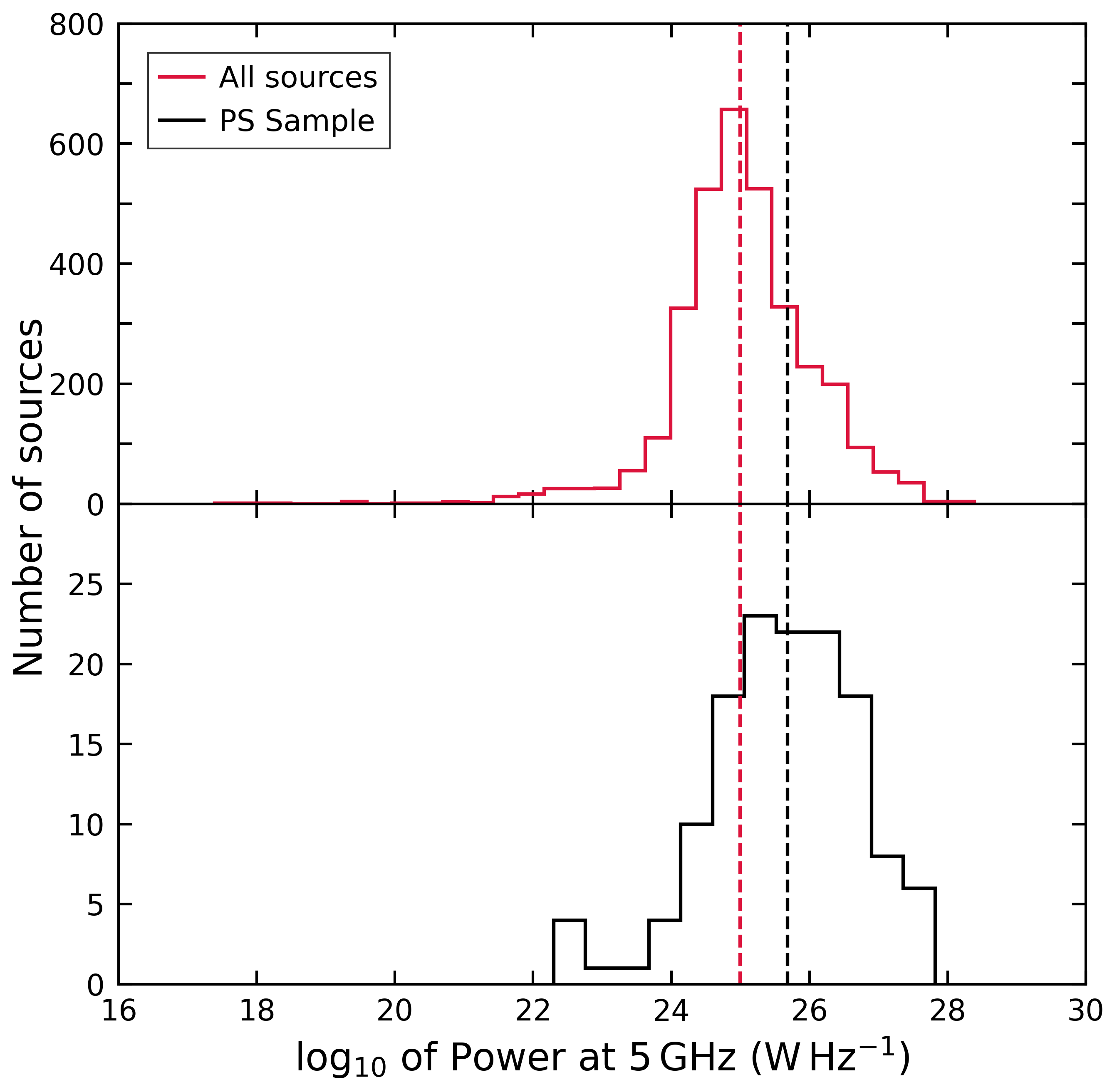}}
    \caption{The distribution of the 5\,GHz radio power for the 3,303 sources from the Master sample that have redshift information available is presented in red in the upper histogram. The black histogram in the bottom plot shows the distribution of the 5\,GHz radio power of the 138 sources from the PS sample for which redshift data is available. The median redshifts of these distributions are plotted as dashed lines in their respective colours. For the total sample, the median \Pfive value and the range of the 16th and 84th percentiles of the distribution is given by $25.0^{+1.0}_{-0.7}$\, $\log_{10}$\,W\,Hz$^{-1}$, and for the PS source sample this is given by $25.7^{+0.9}_{-1.1}$\, $\log_{10}$\,W\,Hz$^{-1}$.}
    \label{fig:pow_dist_master_samp}
\end{figure}

The distribution of the 5\,GHz radio power for the PS sample and the Master sample is provided in Fig. \ref{fig:pow_dist_master_samp}. We find that our PS sample has a median \Pfive\ value that is 10$^{0.7}$ \,W\,Hz$^{-1}$ higher than the median \Pfive\ value of the Master sample. The same is found for the 90\% complete sub-samples of the Master sample and the PS sample, which are not limited by incompleteness of LoLSS. This difference in luminosities between the two samples is likely due to the fact that for a given redshift and flux density in LoLSS, a PS source will have a higher flux density in LoTSS than a simple power-law spectrum source due to its spectral shape. We are thus selecting relatively high flux-density sources in the PS sample, corresponding to the higher average 5\,GHz radio power for PS sources.

\begin{figure}[t]
    \centering
    \resizebox{\hsize}{!}{\includegraphics{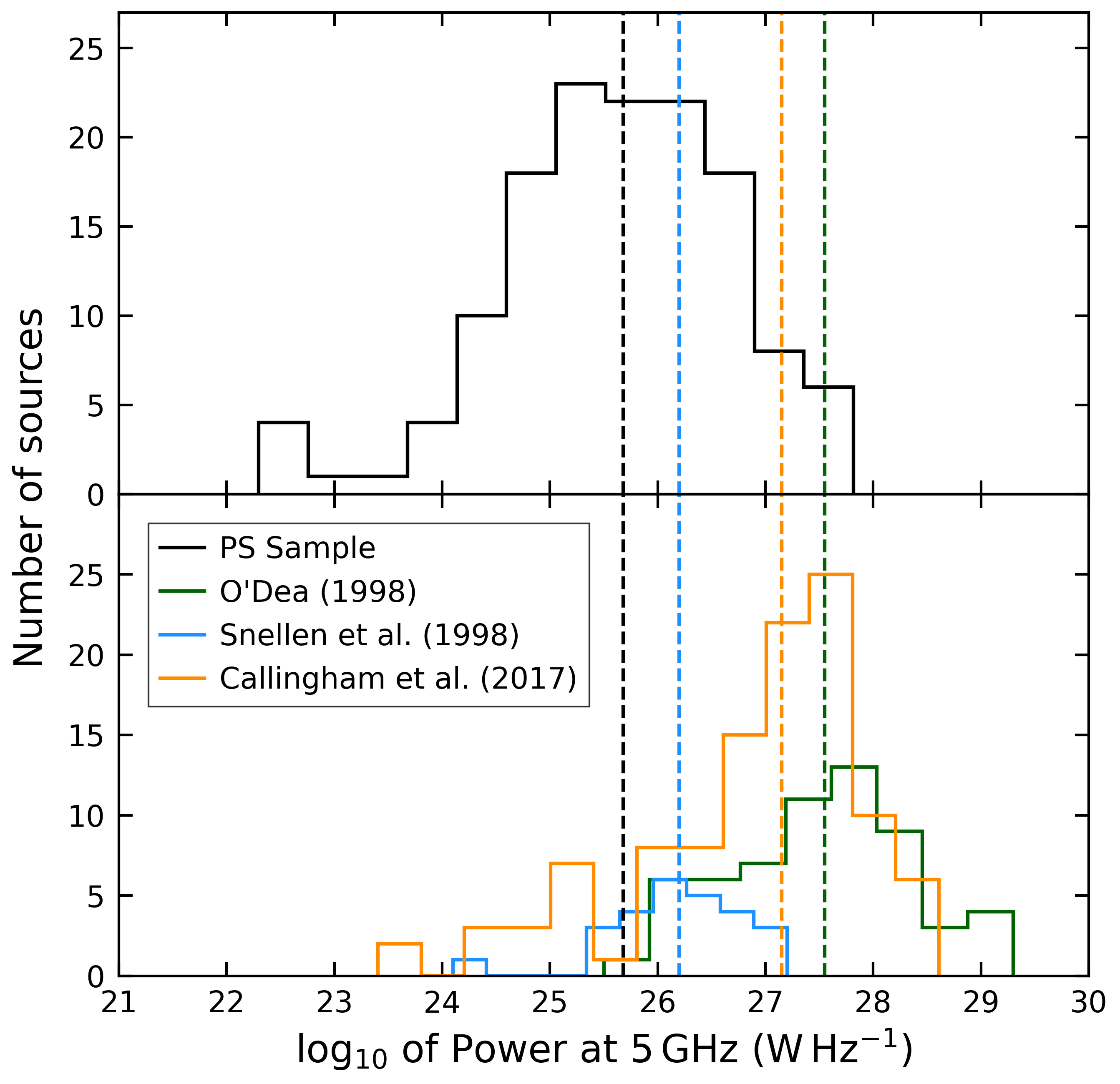}}
    \caption{The distribution of the 5\,GHz radio power for the 138 sources from the PS sample for which redshift data is available is presented in black in the upper histogram. The green, blue, and orange histograms in the lower plot represent the 5\,GHz radio power of the \citet{1998A&AS..131..435S} PS sample, the \citet{O_Dea_1998} PS sample, and the \citet{Callingham_2017} PS sample, respectively. The median redshifts of all distributions are plotted as dashed lines in their respective colours, over both the upper and lower plots. The median $P_{5\,\text{GHz}}$ value and the range of the 16th and 84th percentiles of the distributions of the samples in this study, \citet{1998A&AS..131..435S}, \citet{O_Dea_1998}, and \citet{Callingham_2017} are given by $25.7^{+0.9}_{-1.1}$, $26.2^{+0.5}_{-0.4}$, $27.6^{+0.7}_{-1.1}$, $27.2^{+0.6}_{-1.2}$  ($\log_{10}$\,W\,Hz$^{-1}$), respectively.}
    \label{fig:pow_dist_PS_samples}
\end{figure}

\begin{figure}[t]
    \centering
     \resizebox{\hsize}{!}{\includegraphics{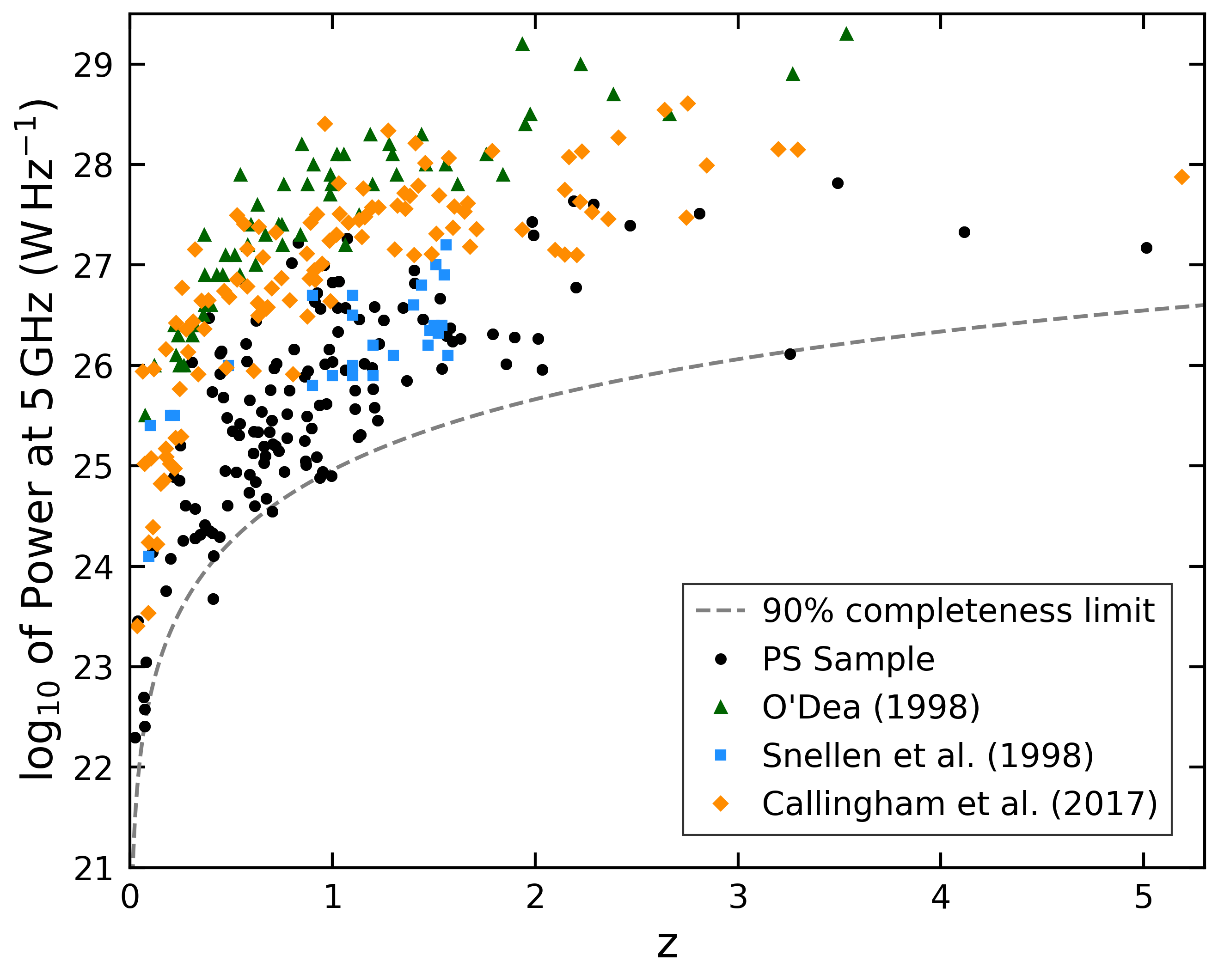}}
    \caption{5\,GHz radio power against redshift for the 138 sources from the PS sample for which redshift data is available is shown as black circles. The green triangles, blue squares, and orange diamonds represent the PS samples of \citet{O_Dea_1998}, \citet{1998A&AS..131..435S}, and \citet{Callingham_2017}, respectively. The dashed grey line corresponds to the 5\,GHz luminosity limit for a source that has a peak flux of 44\,mJy at 144\,MHz, corresponding to the 90\% completeness limit of the PS source selection. To compute this 5\,GHz luminosity limit, a median spectral index of $-0.87$, corresponding to the 90\% limit of \ahigh, was used.}
    \label{fig:z_power}
\end{figure}

In Fig. \ref{fig:pow_dist_PS_samples} we compare the 5\,GHz radio power of our PS sample to the literature PS samples presented by \citet{1998A&AS..131..435S}, \citet{O_Dea_1998}, and \citet{Callingham_2017}. The variations of 5\,GHz radio power with redshift for these samples are shown in Fig. \ref{fig:z_power}.
We note that the 5\,GHz radio powers we find for the sample presented by \citet{Callingham_2017} are higher than those in the original paper, due to a missing factor of four in their radio luminosity calculation. Our PS sample contains the lowest luminosity PS source identified to date with \Pfive $= 2.0 \times 10^{22}$ W Hz$^{-1}$.

In Fig. \ref{fig:LF_z} we also plot the curve for the estimated 90\% limit of luminosity against redshift. To compute this curve, we used the estimated 90\% flux density completeness at 144\,MHz of 44\,mJy described in Section \ref{sec:PS_completeness} and extrapolate this to 5\,GHz using a simple power-law. The spectral index used for this extrapolation (\ahigh = -0.87) corresponds to the 90\% limit of the \ahigh\ distribution for PS sources. We then use this 5\,GHz flux density limit of 2\,mJy to compute the corresponding estimated 90\% limit of the luminosity as a function of redshift.

Compared to the PS samples from \citet{1998A&AS..131..435S}, \citet{O_Dea_1998}, and \citet{Callingham_2017}, the PS sources presented in this study have 5\,GHz radio powers that are, on average, roughly an order of magnitude smaller. Compared to the \citet{O_Dea_1998} and \citet{Callingham_2017} samples, our PS sample does not identify any high luminosity (\Pfive\ $ > 1 \times 10^{28}$ W Hz$^{-1}$) PS sources. However, as can be seen in Fig. \ref{fig:pow_dist_master_samp}, our Master sample also does not contain these high luminosity sources. Therefore, the lack of high luminosity PS sources is likely due to cosmic variance since the LOFAR surveys we use to select PS sources have not yet surveyed the whole sky.

\section{High/low excitation classification of the PS sample}
\label{sec:HERG_LERG}

To investigate the dominant accretion mode for PS sources, we classify sources in our PS sample as `high-excitation' or `low-excitation' radio sources (HERGs and LERGs). HERGs are sources that have a radiatively efficient accretion mode, and radiate strongly across their electromagnetic spectrum \citep[e.g.][and references therein]{2012MNRAS.421.1569B}. LERGs, on the other hand, have an accretion mode that leads to less strong radiative emission throughout the electromagnetic spectrum \citep[e.g.][]{2007MNRAS.376.1849H}. HERGs are shown to be a strongly evolving population, while LERGs show little cosmic evolution \citep{2012MNRAS.421.1569B, 2016MNRAS.460....2P}. Therefore, the classification of PS sources into HERGs and LERGs provides insight into the evolution of the PS source population.

We classified our PS sources into HERGs and LERGs using available optical spectra of our PS sources. We obtained the optical spectra for 54 of our PS sources from the seventeenth data release of the Sloan Digital Sky Survey \citep[SDSS DR17;][]{2022ApJS..259...35A}.

Previous studies have classified sources as HERGs and LERGs based on the relative strength and equivalent width of the 5007\AA\ [\ion{O}{III}] line, with LERGs having significantly less [\ion{O}{III}] emission than HERGs \citep[e.g.][]{1994ASPC...54..201L, 1998MNRAS.298.1035T}. In more recent studies, sources have been classified as HERGs and LERGs based on multiple optical emission lines from separation between Seyfert and LINER galaxies proposed by \citet{2006MNRAS.372..961K} \citep[e.g.][]{2010A&A...519A..48B, 2010A&A...509A...6B}. However, \citet{2010A&A...509A...6B} noted that for a small sample of sources, the use of the Seyfert and LINER diagnostic diagrams can give ambiguous results. We therefore adapt the `excitation index' (EI) classification scheme defined by \citet{2010A&A...509A...6B}.

The EI combines the emission-line ratios of four emission lines, and is defined as EI = $\log_{10}([\ion{O}{III}]$\,/\,H$\beta$) - $\frac{1}{3}$[$\log_{10}([\ion{N}{II}]$\,/\,H$\alpha$) + $\log_{10}([\ion{S}{II}$]\,/\,H$\alpha$) + $\log_{10}([\ion{O}{I}$]\,/\,H$\alpha$)]. The division between HERGs and LERGs is made at a value of EI = 0.95 \citep{2010A&A...509A...6B}, where HERGs have an EI above this limit and LERGs have EIs below this limit. However, the EI classification method requires the presence of all four emission lines. Unfortunately, these four emission lines were not all available for many of the SDSS spectra of the PS sources. Therefore, if a classification based on the EI was not possible, we used the [\ion{O}{III}] emission line only. Here, we identified a source as a HERG when the equivalent width of its [\ion{O}{III}] emission line was above 5\AA, in accordance with selection step (iii) outlined by \citet{2012MNRAS.421.1569B}. 
\begin{table}[t]
    \small
    \centering
    \caption{Breakdown of HERG and LERG classifications}
    \begin{tabular}{l r}
    \hline\hline
    Classification & Number of Sources\\
    \hline
    EI HERG & 8 \\
    EI LERG & 5 \\
    $[$\ion{O}{III}$]$ emission line HERG & 12\\
    \hline
    Total HERG (EI + $[$\ion{O}{III}$]$) & 20\\
    Total LERG & 5\\
    Total Unclassified & 29\\
    \hline
    \textbf{Total} & 54\\
    \hline\hline
    \end{tabular}
    \tablefoot{The details of each classification method are provided in Section \ref{sec:HERG_LERG}.}
    \label{tab:HERG_LERG_selec}
\end{table}

The number of HERGs and LERGs classified using the described classification methods are presented in Table \ref{tab:HERG_LERG_selec}. In total, we can classify roughly half of the PS sources with spectral information available as a HERG or LERG using emission line criteria. For the unclassified sources, not enough emission lines were detected in order to class them. This could be due to a genuine lack of emission lines, suggesting these sources are LERGs. However, the lack of detected emission lines can be due to low signal-to-noise spectra. 

\begin{figure}[t]
    \centering
     \resizebox{\hsize}{!}{\includegraphics{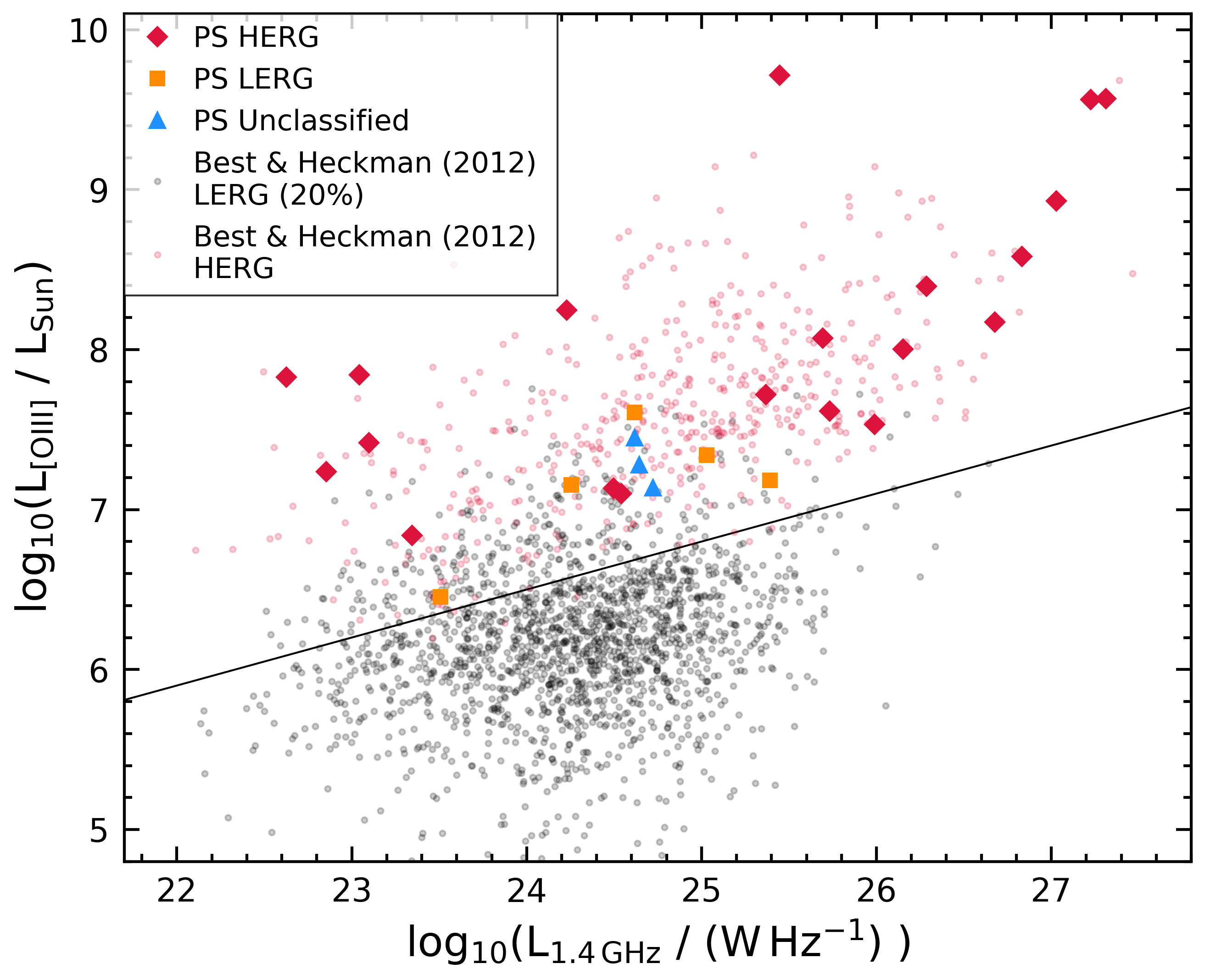}}
    \caption{The [\ion{O}{III}] line luminosity versus 1.4\,GHz radio luminosity of HERG/LERG-classified and unclassified PS sources. The red diamonds, orange squares, and blue triangles indicate PS HERGs, PS LERGs and unclassified PS sources, respectively. The [\ion{O}{III}] line luminosities versus 1.4\,GHz radio luminosities of the sample of HERG and LERG sources presented by \citet{2012MNRAS.421.1569B} are plotted as red and black points, respectively. In order to avoid overcrowding of the figure, only 20\% of the LERGs in the \citet{2012MNRAS.421.1569B} sample are plotted. The solid black line indicates the lower limit to the distribution of HERGs proposed by \citet{2012MNRAS.421.1569B}.}
    \label{fig:OIII_radio_power}
\end{figure}

We therefore investigate the classification mechanism based on the [\ion{O}{III}] line luminosity versus radio luminosity proposed by \citet{2012MNRAS.421.1569B} for the unclassified sources. Figure \ref{fig:OIII_radio_power} shows the distribution of PS sources in the [\ion{O}{III}] line luminosity versus radio luminosity plane. From Fig. \ref{fig:OIII_radio_power}, we find that three unclassified PS sources can be placed in the [\ion{O}{III}] line luminosity versus radio luminosity plane. However, all three unclassified sources lie above the line defined by \citet{2012MNRAS.421.1569B} as the lower limit to the distribution of HERGs. Sources below this line can be classified as LERGs, while sources above this line can be either HERGs or LERGs. This classification mechanism can therefore not definitively class the three unclassified sources into HERGs or LERGs. We note that none of the LERGs in our PS sample lie below the HERG/LERG division line, indicating that the detected PS LERGs have relatively strong [\ion{O}{III}] line emission.

In total, 20 out of the 25 classified PS sources are HERGs, suggesting that our PS sources are more likely to be HERGs than LERGs. This corresponds well to the hypothesised \textit{youth} model for PS sources, as HERGs are shown to be strongly evolving sources at $z \lesssim 1$ \citep{2012MNRAS.421.1569B, 2016MNRAS.460....2P}. Additionally, roughly a third of the PS sources that we were able to classify as HERGs or LERGs have $L_{\text{1.4\,GHz}} \gtrsim 10^{26}$\,W\,Hz$^{-1}$, all of which have been classified as HERGs. This supports previous observations that HERGs dominate the population at high luminosities \citep[e.g.][]{2012MNRAS.421.1569B, 2018A&A...620A..16B}.

In our sample of PS sources we have an overabundance of HERG sources compared to a general sample of AGN, which is dominated by LERGs \citep{2012MNRAS.421.1569B, 2016MNRAS.460....2P}. However, in our characterisation of PS sources we have not taken redshift into account. At high redshifts we expect to find more HERGs since these will have more active spectra and stronger optical emission. This will cause them to have a higher likelihood of being detected by optical surveys. 

To compare the classification of our PS sample without this redshift bias, we apply the redshift restriction $z \leq 0.3$ from the HERG and LERG classification of \citet{2012MNRAS.421.1569B}. In this redshift range, we classify 5 PS sources as EI-LERGs, 6 PS sources as HERGs (5 EI-HERGs, 1 [\ion{O}{III}] HERG), and 1 source cannot be classified. Furthermore, the PS sources that were identified as HERGs in this local sample all have a 1.4\,GHz luminosity $< 10^{25}$\,W\,Hz$^{-1}$. This shows that without a redshift and luminosity bias, we still find a significant overabundance of HERGs in our PS sample, as \citet{2012MNRAS.421.1569B} and \citet{2016MNRAS.460....2P} identify HERGs and LERGs in a roughly one-to-ten ratio in this redshift and luminosity range, instead of the one-to-one ratio for our PS sample. Therefore, the fact that most of our PS sources are classed as HERGs indicates the population is a quickly evolving one, as is consistent with the \textit{youth} model of the radio emission. {However, we note that more PS sources with optical spectral information are needed to confirm if this overabundance is observed for our entire PS sample.}

\section{Euclidean normalised source counts}
\label{sec:source_counts}
In order to investigate the evolution of PS sources, we determined the radio source counts for different samples of PS sources and compared these with the counts predicted by evolutionary models of radio populations. 

\begin{figure}[t]
    \centering
     \resizebox{\hsize}{!}{\includegraphics{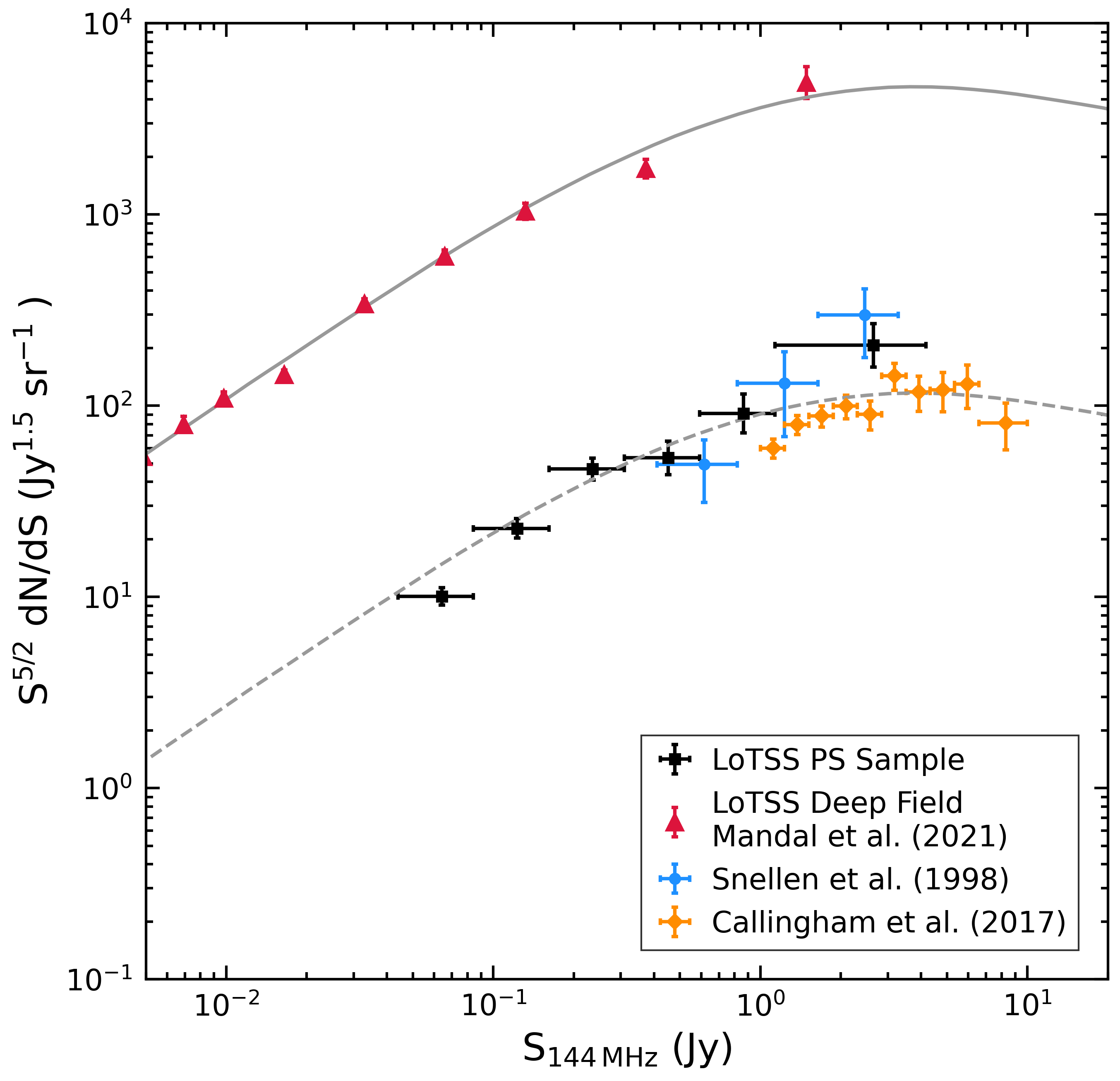}}
    \caption{The Euclidean normalised differential source counts for different samples of PS sources. The black squares show our sample of PS sources. The blue circles show the \citet{1998A&AS..131..435S} PS sample converted to 144\,MHz. The orange diamonds show the \citet{Callingham_2017} PS sample. For reference, the source counts for the LoTSS Deep Field derived by \citet{2021A&A...648A...5M} are plotted as red triangles. The solid grey line shows the model for 150\,MHz AGN source counts proposed by \citet{2010MNRAS.404..532M}. The dashed grey line shows the \citet{2010MNRAS.404..532M} model scaled down by a factor 40, which is consistent with the PS source data.}
    \label{fig:source_counts}
\end{figure}

The differential source counts, $dN/dS$, were computed for our sample of PS sources with 144\,MHz flux densities $\geq 44$\,mJy, corresponding to the flux density completeness limit of our sample. $dN/dS$ was computed by summing the observed number of PS sources in {six} 144\,MHz flux bins and dividing these by the detected area of the sky, $A$. In the case of our PS sample, we used the detection area of LoLSS \citep[$A = 740$ deg$^2$;][]{2021A&A...648A.104D}, as this is the survey that limits the detection area for our PS sources. We then normalised the source counts with a factor $S^{2.5}$, which corresponds to normalising to a uniformly distributed Euclidean space. The resulting normalised differential source counts are shown in Fig. \ref{fig:source_counts} and listed in Table \ref{tab:source_counts}.

\begin{table}[t]
    \small
    \centering
    \caption{144\,MHz normalised differential radio source counts for our sample of PS sources}
    \begin{tabular}{c c c || c c c }
    \hline\hline
    \multicolumn{3}{c||}{\textbf{LoTSS}} & \multicolumn{3}{c}{\textbf{GLEAM}} \\
    {$\langle S \rangle$ [Jy]} & $N_S$ & $N^{+\sigma}_{-\sigma}$ {[Jy$^{1.5}$/\,sr]} & $\langle S \rangle$ {[Jy]} & $N_S$ & $N^{+\sigma}_{-\sigma}$ {[Jy$^{1.5}$/\,sr]}\\
    \hline
    0.06 & 101 & $ 10.06 ^{+ 1.10 }_{- 1.00 }$ & 1.12 & 78 & $ 60.09 ^{+ 7.60 }_{- 6.79 }$\\
    0.12 & 81 & $ 22.83 ^{+ 2.83 }_{- 2.53 }$  & 1.38 & 77 &$ 79.74 ^{+ 10.16 }_{- 9.07 }$\\
    0.24 & 63 & $ 46.68 ^{+ 6.65 }_{- 5.86 }$  & 1.70 & 62 &$ 88.41 ^{+ 12.71 }_{- 11.19 }$\\
    0.45 & 29 & $ 53.40 ^{+ 11.86 }_{- 9.85 }$ & 2.09 & 49 &$ 99.56 ^{+ 16.35 }_{- 14.17 }$\\
    0.86 & 22 & $ 91.34 ^{+ 23.90 }_{- 19.31}$ & 2.58 & 33 &$ 90.21 ^{+ 18.59 }_{- 15.62 }$\\
    2.65 & 18 & $ 207.4 ^{+ 61.3 }_{- 48.4 }$  & 3.18 & 38 &$ 143.6 ^{+ 27.3 }_{- 23.1 }$\\
    & &                                        & 3.92 & 23 &$ 118.0 ^{+ 30.1 }_{- 24.4 }$\\
    & &                                        & 4.83 & 18 &$ 121.4 ^{+ 35.9 }_{- 28.3 }$\\
    & &                                        & 5.96 & 15 &$ 130.2 ^{+ 43.0 }_{- 33.2}$\\
    & &                                        & 8.29 & 13 &$ 81.10 ^{+ 29.27 }_{- 22.18}$\\

    \hline\hline
    \end{tabular}
    \tablefoot{$\langle S \rangle$ is the centre of the respective flux bin in Jy, {$N_S$ corresponds to the number of sources in each flux density bin,} and $N$ gives the normalised differential source counts, and $\pm\sigma$ are the Poissonian errors on the source counts. {The LoTSS and GLEAM columns indicate whether the source counts correspond to the LoTSS PS sample presented in this study or the GLEAM PS sample presented by \citet{Callingham_2017}.} We only report the source counts of the \citet{Callingham_2017} above their $\approx$100\% completeness limit.}
    \label{tab:source_counts}
\end{table}

We can use this same method to construct the normalised radio source counts for the PS sample presented by \citet{Callingham_2017}. To construct the source counts at 144\,MHz for this sample, we used the reported GLEAM flux densities at 143\,MHz, and assumed the flux density does not change significantly over 1\,MHz. Only sources with 143\,MHz flux densities $\geq 1$\,Jy, corresponding to the estimated 100\% completeness limit of this PS sample, were considered when evaluating these source counts. The computed source counts for the \citet{Callingham_2017} PS sample are also shown in Fig. \ref{fig:source_counts} and listed in Table \ref{tab:source_counts}.

We also compare the radio source counts for our sample of PS sources to the radio source counts derived by \citet{1998A&AS..131..435S} for their sample of PS sources. These source counts were evaluated at the peak frequency of the individual PS sources, which was assumed to correspond to a median frequency of 2\,GHz. We have therefore determined the radio source counts for this sample at 144\,MHz by shifting the counts from 2\,GHz to 144\,MHz assuming the power-law model from Equation \eqref{eq:simple_pow_law} with a spectral index of -0.80. The resulting source counts are shown in Fig. \ref{fig:source_counts}. Note that we have excluded the highest flux density bin since this was not derived directly from the sample presented by \citet{1998A&AS..131..435S}, and likely contains uncharacterisable systematics.

In order to compare the radio source counts for the different PS samples to a general sample of AGN, the 150\,MHz LoTSS Deep Field radio source counts determined by \citet{2021A&A...648A...5M} are also plotted in Fig. \ref{fig:source_counts}, along with the 150\,MHz model for AGN source counts presented by \citet{2010MNRAS.404..532M}. Moving this model down by a factor 40 roughly agrees with the observed curve for the different observed PS source counts. 

The interpretation of this factor of 40 between the total and PS source counts needs to be treated with caution. While it is tempting to assume that if PS sources evolve into the larger scale sources sampled by \citet{2021A&A...648A...5M}, they should undergo similar cosmological evolution since their lifetime is significantly less than Hubble time. Therefore, this factor of 40 would encode the ratio of the lifetime of the two source classes and the luminosity function of PS sources. If taken at face value, this factor implies that the lifetime of the PS phase is $\approx$40 times shorter than the lifetime of a large radio galaxy at low-frequencies.

\citet{1998A&AS..131..435S} found that the source counts at 2\,GHz for their PS sample are scaled down by a factor of 250 compared to the source counts of large-scale radio galaxies. If we assume this factor also encodes the lifetime of this sample of PS sources, that would mean that PS sources selected at 2\,GHz have much shorter lifetimes, and thus stronger evolution, than those selected at 144\,MHz. This stronger evolution could be related to the jet power of the sources, since at 2\,GHz the detected radiation is much closer to the core than at 144\,MHz, causing the detected jet power to be stronger too. Closer to the core, evolution of the galaxy would then take place more quickly than in the outer regions. However, this interpretation of the scaling factor does not take into account any redshift evolution, and is probably too simplistic as the redshift evolution of AGN and PS sources are expected not to be identical \citep{2007A&A...463...97L, Kunert_Baj/j.1365-2966.2010.17271.x}. To decouple the impact of any potential luminosity evolution from the source counts, we investigate the luminosity function in the following section.

\section{Peaked-spectrum radio luminosity function}
\label{sec:LFs}
The second method we used to characterise the evolution of our PS sources is the luminosity function. We compute the luminosity function for both our Master sample and our PS sample at 144\,MHz using the standard 1/\Vmax\ method \citep{1968ApJ...151..393S}, with the luminosity function in a given luminosity bin centered at $L$ given by:
\begin{align}
    \label{eq:LF}
    \frac{dN(L_j)}{d \log L} &= \frac{1}{\Delta \log L_j} \sum_{i=1}^{N}, \frac{1}{V_{i}}
\end{align}
where the sum is over all $N$ sources in the luminosity bin. $V_{i}$ corresponds to the volume over which a galaxy can be detected given the optical and radio selection criteria for the sample. This is calculated as $V_{i} = V_{\text{max}} - V_{\text{min}}$, where $V_{\text{max}}$ and $V_{\text{min}}$ are the volumes corresponding to the upper and lower redshift limits, respectively, for which a source could be detected in our sample. Below, we outline our methods of estimating our selection criteria that determine our optical and radio completeness, and \Vmax.

\subsection{Estimating selection effects}
\label{sec:det_lim}
In order to calculate the maximum volume over which a galaxy can be detected, both the radio and optical detection limits need to be characterised. For the radio detection limit of PS sources we can use the flux density limit of $S_{\text{144\,MHz, PS lim}} = 44$\,mJy described in Section \ref{sec:PS_completeness}. However, since the spectral shape of a source determines the extrapolated flux density at 144\,MHz, we have to use a different spectral index for the extrapolation for the Master sample and the PS sample. For the flux extrapolation of the Master sample we use a spectral index \alow\ = -1.1, corresponding to the lower 95\% limit to the distribution of \alow\ of the Master sample. This results in a limiting flux density for the Master sample of $S_{\text{144\,MHz, lim}} = 13$\,mJy.

Besides the radio survey selection limits, we also need to take the optical selection limits into account. However, the LoTSS optical catalogues are a combination of different optical surveys, and thus have no defined detection limit. In order to define a consistent optical detection limit, we only use sources from our Master sample that have counterparts in {SDSS-DR7} \citep{2009ApJS..182..543A}. These counterparts were obtained by crossmatching the position of SDSS sources with the positions of LoTSS sources using a crossmatching radius of 2\arcsec. To compute the optical \Vmax, we use the SDSS \textit{g-} and \textit{i-}band photometry -- implying we only consider sources for which both the SDSS \textit{g-} and \textit{i-}band magnitudes are measured. In total this Master sub-sample with SDSS counterparts consists of 1,909 Master sample sources, of which 105 are PS sources. We then set the optical detection limit as the 95\% completeness-limit of SDSS in the \textit{i-}band ($ m_{i, \text{lim}} = 21.3$ mag). {We refer to this sample as the SDSS-selected sample.}

{To test the influence of the optical incompleteness on the resulting luminosity function, we also compute the luminosity functions using the Pan-STARRS \textit{g-} and \textit{i-}band photometry included in the LoTSS DR1 optical catalogue \citep{2019A&A...622A...2W}. To ensure uniform selection effects for this sample, we use only the photometric redshifts. We refer to this sample as the photo-z selected sample. In total, this results in a Master sub-sample of 2,156 sources, of which 88 sources are PS sources. We set the detection limit of this sample as a conservative estimation for where the photometric redshifts are still well-calibrated ($ m_{i, \text{lim}} = 22.5$ mag).}

\subsection{Estimating \Vmax}
Using the detection limits described in Section \ref{sec:det_lim}, we can calculate the maximum volume in which a source can be detected by our optical and radio surveys, $V_{\text{max, opt}}$ and $V_{\text{max, radio}}$, respectively. In order to calculate $V_{\text{max, opt}}$, we first calculate the absolute \textit{i}-band magnitude, $M_i$, for each source as:
\begin{align}
\label{eq:abs_i}
    M_i = m_i - \text{DM} - K_i(z),
\end{align}
where $m_i$ is the apparent magnitude, DM is the distance modulus, and $K_i(z)$ is the \textit{k}-correction. The \textit{k}-corrections are calculated with the \texttt{K-corrections calculator} \citep{2010MNRAS.405.1409C, 2012MNRAS.419.1727C} for the \textit{g-i} colour. To determine $V_{\text{max, opt}}$, we evaluate Equation \ref{eq:abs_i} at $m_i =  m_{i, \text{lim}}$, at a series of redshifts ($\Delta z = 0.0001$) to find the greatest redshift, $z_{\text{max, optical}}$, at which $M_{i, \text{lim}} \leq M_{i, \text{source}}$. Above this redshift, the source is too faint to be detected by {the optical survey}. 

Analogously, the maximum radio redshift, $z_{\text{max, radio}}$, is found by first computing the 144\,MHz radio luminosity, $L_{\text{144\,MHz}}$ of each source using Equation \eqref{eq:P5GHz}, evaluated at 144\,MHz instead of 5\,GHz. We then compute Equation \eqref{eq:P5GHz} at $S_{{\text{144\,MHz}}} = S_{\text{144\,MHz, (PS) lim}}$ at the same series of redshifts as the optical method. From this we can identify $z_{\text{max, radio}}$ as the smallest redshift where $L_{\text{144\,MHz, (PS) lim}} \geq L_{\text{144\,MHz, source}}$. Above this redshift, the source is too faint to be detected in our sample.

The combined maximum optical and radio redshift limits for each source $z_{\text{max}}$ is now the minimum of $z_{\text{max, radio}}$ and $z_{\text{max, opt}}$. We then calculate \Vmax\ from the integrated comoving volume corresponding to $z_{\text{max}}$ and multiplying this volume by the fraction of the sky covered by our sample. This area of detection corresponds to the fractional sky coverage of LoLSS ($740$ deg$^2$), which is the survey that limits our detection area.

\subsection{Computing the radio luminosity function and implications}
\begin{figure*}[ht]
    \centering
    \includegraphics[width=17cm]{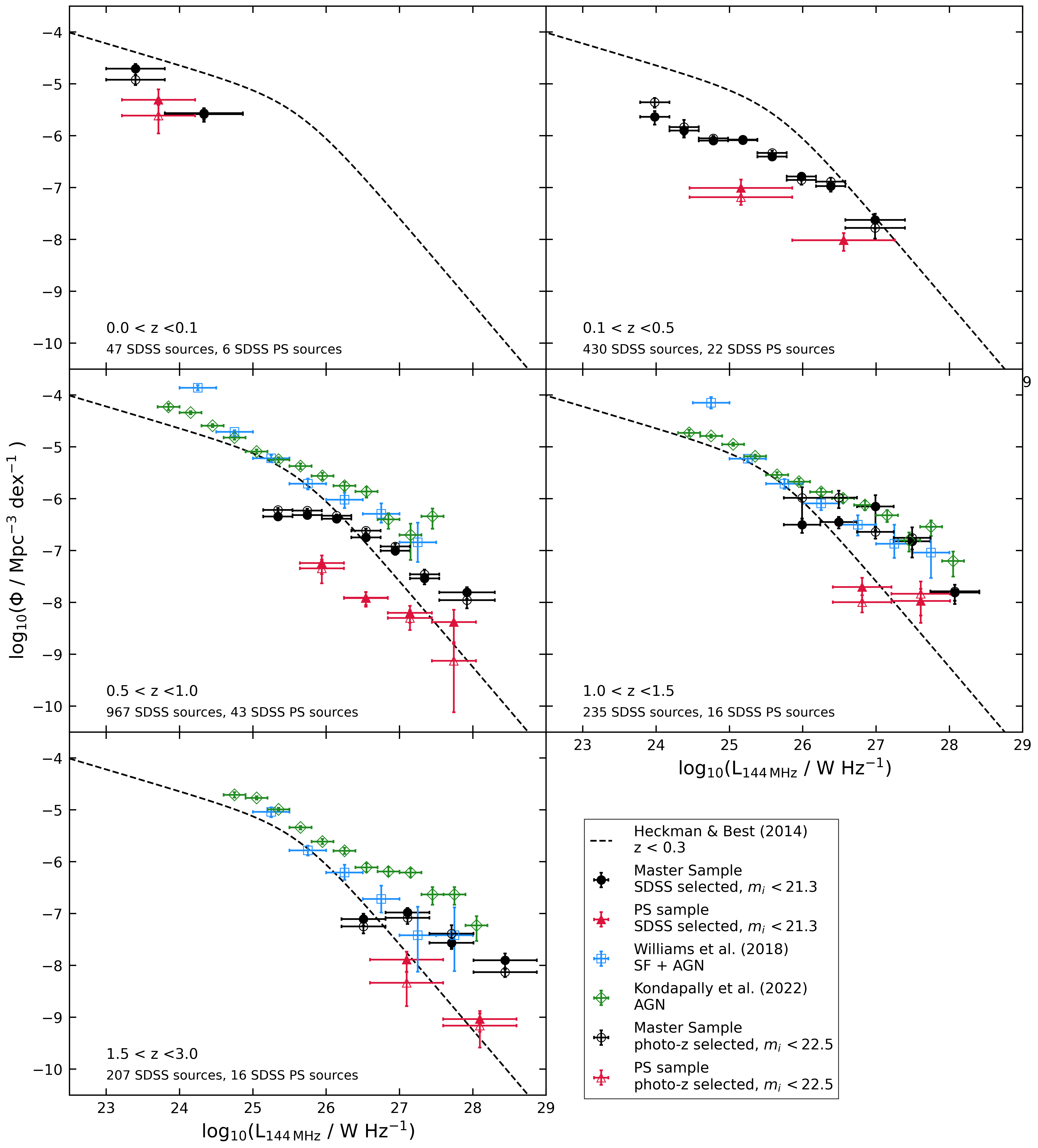}
    \caption{The 144\,MHz luminosity function, $\Phi$, for the Master Sample (black circles) and PS sample (red squares) at different redshifts. The filled symbols show the SDSS selected sample, and the open symbols show the photo-z selected sample.    The number of sources in the Master sample and PS sample for each redshift range are shown in the bottom-left of their respective plots. The grey curve corresponds to the double power-law model for the AGN local ($z \simeq 0.1$) luminosity function at 1.4\,GHz parameterised by \citet{2014ARA&A..52..589H}, converted to 144\,MHz using a spectral index $-0.7$. {The 150\,MHz luminosity functions for the SF+AGN sample presented by \citet{2018MNRAS.475.3429W}, and the AGN sample presented by \citet{2022MNRAS.513.3742K} are shown as blue squares, and green diamonds, respectively. We note that for the $z > 1$ bins, both the PS and Master sample luminosity functions are not representative of a complete sample due to their optical incompleteness. However, since the optical incompleteness is identical between the PS and Master samples, the offset is an true reflection of evolution between the samples.}}
    \label{fig:LF_z}
\end{figure*}
We compute the luminosity function using Equation \eqref{eq:LF} for sources in five different redshift ranges to investigate the evolution of our PS sample. The edges of these redshift ranges were chosen as 0.1, 0.5, 1.0, 1.5 and 3.0, to allow for a sufficient number of sources in each redshift range, while being small enough to probe any evolution in the luminosity functions between redshift ranges. These redshift limits correspond to the lower redshifts for each respective redshift bin used to calculate $V_{\text{min}}$. We do not evaluate the luminosity function for luminosities below $10^{23}$\,W\,Hz$^{-1}$, since star-forming galaxies (SFGs) dominate the luminosity function at these low luminosities \citep[e.g.][]{2019A&A...622A..17S,2021PASA...38...41F}, while for this study we are only interested in the AGN population. 

We compute the luminosity function for each redshift range and for both the Master sample and the PS sample. The size of the luminosity bins, $\Delta \log L$, is defined separately for the Master and PS sample in each redshift bin (see Table \ref{tab:LF_z}) in order to have the most robust and uniform number of sources as possible in all luminosity bins. {For easy comparison, the same luminosity bins are used for the SDSS and photo-z selected samples.} We note that we are limited in our analysis by the small number of PS sources in all redshift ranges. This is most notable in the `local' redshift bin ($z <$ 0.1) where the PS source luminosity function could only be evaluated in one luminosity bin. 

We estimate the Poissonian counting error on ${dN(L_j)}/{d \log L}$ as:
\begin{align}
    \sigma_j &= \left[\frac{1}{(\Delta \log L_j)^2} \,\sum_{i=1}^{N} \frac{1}{V^2_{i}}\right]^{1/2}.
\end{align}
In luminosity bins with a small number of sources ($N<5$), we use the 84\% upper and lower confidence limits estimated from Poisson statistics, using Equations (9) and (12) of \citet{1986ApJ...303..336G}. Our resulting luminosity functions are shown in Fig. \ref{fig:LF_z}, and tabulated in Appendix \ref{ap:LF}.

{The luminosity functions for the SDSS and photo-z selected samples in Figure\,\ref{fig:LF_z} agree within $\sim$1$\sigma$ throughout redshift and luminosity space -- implying it is the radio incompleteness that dominates the calculation for \Vmax. However, both optical samples only include roughly a third of our PS sample, with the rest of the sample likely being too faint for optical detection. Optical incompleteness will impact the luminosity functions at high redshifts, ensuring that the luminosity functions at $z > 1$ do not represent a complete sample. However, since the optical characteristics of our PS sample and a general sample of radio-loud AGN are similar \citep{2007A&A...463...97L, 2022MNRAS.511..214N} we can assume that this optical incompleteness will be similar for the PS and Master sample. We conclude that the observed differences between these two samples will not be dominated by selection effects.}

In Fig. \ref{fig:LF_z} we also plot the double power-law model for the local ($z \simeq 0.1$) AGN luminosity function derived at 1.4\,GHz by \citet{2014ARA&A..52..589H} but shifted to 144\,MHz by using a spectral index of $-$0.7. {We plot the \citet{2014ARA&A..52..589H} model for all redshift ranges to help guide the readers eye.} For a complete sample of AGN, the measured luminosity functions in our `local' redshift range are expected to line up with this model, as was shown by \citet{2019A&A...622A..17S} for a sample of AGN in LoTSS-DR1. However, as can be seen in the two upper plots of Fig. \ref{fig:LF_z}, the local luminosity functions of both the Master sample and PS sample do not line up with this model, but are shifted down by a factor of $\sim$4.

{For the $z > 0.5$ redshift bins, we plot the \citet{2018MNRAS.475.3429W} SF+AGN, and the \citet{2022MNRAS.513.3742K} AGN luminosity functions at 150\,MHz. For the \citet{2022MNRAS.513.3742K} sample, the redshift bins do not align with the redshift bins in this work, so we have used the redshift ranges $0.7 < z < 1.0$, $1.3 < z < 1.7$, and $1.7 < z < 2.1$. For the \citet{2018MNRAS.475.3429W} sample, the edges of the redshift bins are 0.50, 1.00, 1.50, and 2.00.}

{In the $0.5<z<1.0$ redshift bin, we also find an offset of a factor $\sim$4 between our Master sample luminosity function and the luminosity functions presented by \citet{2018MNRAS.475.3429W} and \citet{2022MNRAS.513.3742K}.} This offset is present because our first two selection criteria for the Master sample (the source has to be isolated in LoTSS, and the source has to be unresolved in LoTSS), have not been corrected for when computing the luminosity function. We therefore only use a subset of the total AGN population in our detection area to compute the number densities, causing the total luminosity function to be shifted down. {However, we note that} in the first redshift bin ($z < 0.1$) our ability to compare the Master sample luminosity function to the model is limited by the small number of sources available ($N = 47$ for the Master sample, $N = 6$ for the PS sample).

{However, in the two highest redshift bins ($z > 1.0$) the offset between the Master sample and literature samples luminosity functions is not present. This can be explained from the fact that at high redshifts most sources will be unresolved. This means that at high redshifts, the Master sample will be close to a complete sample, causing the luminosity function to align with the literature samples.}

In the four highest ($z \geq 0.1$) redshift ranges, the Master sample luminosity functions flatten and even displays a turnover at low luminosities for the redshift ranges $>$0.5. This feature is not observed in previous studies of the 150\,MHz luminosity function for complete AGN samples at high redshifts \citep{2021A&A...656A..48B, 2022MNRAS.513.3742K}. We conclude that the turnover at low luminosities in our sample is caused by incompleteness, mostly because of our optical surveys.

\begin{figure}[t]
    \centering
    \resizebox{\hsize}{!}{\includegraphics{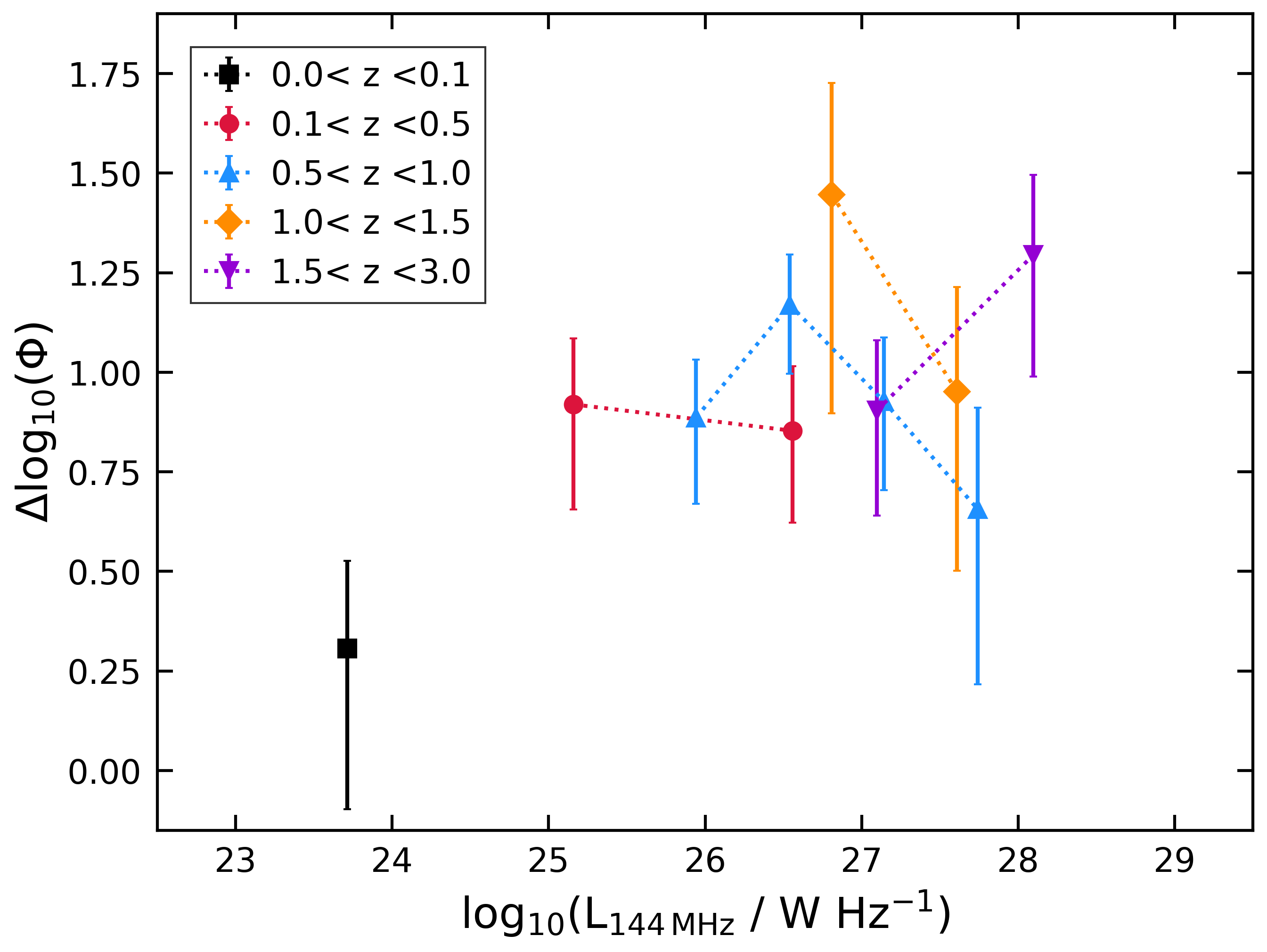}}
    \caption{The difference between the Master sample luminosity function and the PS sample luminosity function, at each PS sample luminosity bin for different redshift ranges.}
    \label{fig:LF_resid}
\end{figure}

However, as this incompleteness is due to survey and selection effects, it is similar for both the Master and PS samples. Therefore, we can use the relative offset between the luminosity functions of the Master sample and the PS sample in each redshift range to estimate the cosmological evolution of our PS sample. These offsets, $\Delta \log(\Phi)$, were found by interpolating the Master sample luminosity functions to the centres of the PS sample luminosity bins, and computing $\Delta \log(\Phi) = \log(\Phi_{\text{Master}}) -  \log(\Phi_{\text{PS}})$. Here $\Phi_{\text{Master}}$ and $\Phi_{\text{PS}}$ are the luminosity functions at the centres of the PS sample luminosity bins for the Master sample and PS sample, respectively. The errors on $\Delta \log(\Phi)$ were computed as the quadratic sum of the Poissonian counting errors of the two luminosity functions. The resulting offsets between the Master and PS sample are shown {for the SDSS selected sample} in Fig. \ref{fig:LF_resid}, and tabulated in Table \ref{tab:LF_z}. {The photo-z selected offsets are not shown in Fig. \ref{fig:LF_resid}, since the difference in the relative offsets for the two different samples is negligible.}

Figures \ref{fig:LF_z} and \ref{fig:LF_resid} show that for the lowest redshift range ($z < 0.1$), the offset of the PS sample luminosity function to the Master sample luminosity function is smaller than for the four highest redshift ranges. In this lowest redshift bin, the offset between the two luminosity functions is close to zero, whereas at higher redshifts the offset is about one order of magnitude. Such a result is somewhat surprising as it has been generally assumed that the local Universe is underdense in PS sources \citep[e.g.][]{Snellen_2000,2007A&A...463...97L, Kunert_Baj/j.1365-2966.2010.17271.x}. However, since we only identified 6 PS sources with $z < 0.1$, caution needs to be taken in over-interpreting such data.

More importantly, the offset for all four redshift ranges with $z > 0.1$ is similar, with the PS luminosity function shifted down by a factor $\sim$10 compared to the Master sample. As all these redshift ranges only probe sources with $L_{\text{144\,MHz,}} > 10^{25}$\,W\,Hz$^{-1}$, this informs us that there are approximately 10 Master sample sources for every one PS source in a given volume at $L_{\text{144\,MHz,}} > 10^{25}$\,W\,Hz$^{-1}$.

{From Figure \ref{fig:LF_z}, we conclude that the Master Sample luminosity function is a factor $\sim$4 smaller than that of a complete sample \citep{2018MNRAS.475.3429W, 2022MNRAS.513.3742K} in the redshift ranges $z < 1.0$. From this, we can conclude that there are $\sim$40 radio-loud AGN sources for every PS source for $L_{\text{144\,MHz,}} > 10^{25}$\,W\,Hz$^{-1}$ and $z<1.0$. For $z > 1.0$ this offset reduces to a factor $\sim$10 between a complete AGN sample and PS sample.}

If we assume the \textit{youth} scenario for PS sources is correct, PS sources will evolve into Master sample sources on smaller timescales than the redshift ranges probed. Indeed, because we observe similar offsets between the Master and PS luminosity functions for $L_{\text{144\,MHz,}} > 10^{25}$\,W\,Hz$^{-1}$ at all redshifts ranges, we can conclude that the birth- and death-rate of PS sources stays the same between redshifts of 0.1 to 3. This confirms that PS source lifetimes are short compared to cosmological time scales.

The shape of the PS luminosity function also remains approximately constant relative to the Master sample and to the {complete} AGN population{s presented by \citet{2018MNRAS.475.3429W} and \citet{2022MNRAS.513.3742K}.}. A similar luminosity function for PS sources and large scale radio-loud AGN implies PS sources evolve into large-scale radio sources as their cosmological evolution is the same. Such a result lends strong support to the \textit{youth} model for PS sources. A constant birth-rate for PS sources across $z > 0.1$ also implies that {a fraction of} radio-loud AGN also fade to the point that they will no longer be identified as high ($L_{\text{144\,MHz,}} > 10^{25}$\,W\,Hz$^{-1}$) luminosity. If this was not the case, there would be an overabundance of radio-loud AGN in the local Universe relative to the distant Universe.

However, we note that the offset between the Master and PS sample luminosity functions is less significant for sources with $L_{\text{144\,MHz,}} < 10^{25}$\,W\,Hz$^{-1}$. This suggests that there are differences between PS sources with low luminosities and those with high luminosities. It is possible that at these low luminosities we are potentially also detecting \textit{frustrated} PS sources, which have such a low jet power that the jet may not ever penetrate the interstellar medium of its host galaxy. In this case, the birth- and death- rates of the young population of PS sources would be the same as in the more distant universe, but the local Universe has an additional number of low-luminosity, \textit{frustrated} PS sources that we are not sensitive to at high redshifts in our flux density-limited surveys. However, additional detections over a larger detection area of local, low luminosity ($L_{\text{144\,MHz}} < 10^{25}$\,W\,Hz$^{-1}$) PS sources are necessary to confirm this hypothesis.

\section{Conclusions}
\label{sec:conclusion}
In this study, we have identified 373 PS sources with spectral peaks around $144$\,MHz using the LOFAR-surveys LoTSS, and LoLSS, as well as NVSS. A source was identified as a PS source when the power-law fit between its LoLSS and LoTSS flux density measurements was positive, while the power-law between the LoTSS and NVSS flux densities was negative. For comparison purposes, we also defined a Master sample of unresolved radio sources that made no assumption about the spectral shape of the source, and is the sample from which the PS sample was drawn.

Our PS sample has increased the number of known PS sources around the HETDEX area by a factor 50. Using the LoTSS ancillary optical catalogue, we are able to identify the redshift for 138 out of 373 PS sources, of which 54 were spectroscopic redshifts. The 5\,GHz radio luminosity distribution computed for our PS sample shows that we have identified the lowest average radio power PS sample to date by roughly an order of magnitude.

Using this sample, we investigated the evolution of our PS sample using HERG and LERG classification, source counts, and luminosity functions. We found the following:
\begin{itemize}
    \item \textbf{HERG/LERG classification:} Using optical line emission criteria we were able to identify 25 PS sources as HERGs (20 sources) or LERGs (5 sources). When we take redshift bias into account, we find a one-to-one ratio of HERGs and LERGs in our PS sample, showing our PS sources are predominantly HERGs compared to a general AGN population. Since HERGs are a quickly evolving population, this suggests our PS sample will be dominated by quickly evolving sources.
    \item \textbf{Source counts:} We evaluated the Euclidean normalised source counts at 144\,MHz for our PS sample, as well as two known PS samples \citep{1998A&AS..131..435S,Callingham_2017}. When compared to the source counts for a general sample of AGN, the PS source counts are scaled down by a factor of 40. From this we conclude that the lifetime of PS sources is about 40 times shorter than that of low-frequency radio-loud AGN, assuming the luminosity function of both populations are the same.
    \item \textbf{Luminosity function:} With the redshift information available, we computed the radio luminosity function for our PS sample. This is the the first time a PS luminosity function has been produced for a homogeneously-selected PS sample. We demonstrate that for redshifts $>$ 0.1, and sources with $L_{\text{144\,MHz}} \gtrsim 10^{25}$\,W\,Hz$^{-1}$, {the offset between the PS luminosity function and that of the Master sample remains constant}. We interpret this as strong evidence that these high luminosity PS sources evolve into large-scale radio-loud AGN. {Such a conclusion also implies that there is one PS source for every $\sim$10 unresolved high-luminosity radio-loud AGN, and the rate at which PS sources enter the later population is roughly consistent with the rate at which the large AGN fade to lower luminosities.} If this was not the case, there would be an overabundance of PS sources relative to their evolved counterparts. However, in the local ($z < 0.1$) Universe, we note that the offset between the PS and Master sample luminosities is smaller at luminosities less than $10^{25}$\,W\,Hz$^{-1}$. We interpret this as the potential addition of frustrated sources in the population, which do not have the power to evolve into large-scale radio-loud AGN. We do not see this population at higher redshifts as our flux-limited surveys are not sensitive to these low-luminosity sources at those distances.
\end{itemize}

We conclude that our HERG/LERG classification, source counts, and luminosity function analyses all indicate that our population of PS sources is a quickly evolving radio-loud AGN population. We suggest this provides strong support that the \textit{youth} scenario applies to the majority of our PS sources with low-frequency luminosities $\gtrsim 10^{25}$\,W\,Hz$^{-1}$. Besides, the relative lifetimes and abundances of PS sources found using the source counts and luminosity functions are in agreement with each other, further supporting this hypothesis. However, we do note that the surplus of PS sources at luminosities lower than this still allows the \textit{frustrated} hypothesis of PS sources to apply.

To test if these sources are in fact frustrated, precise broadband spectral modelling is required. Furthermore, a larger sample of PS samples, as will derivable once the LoLSS and LoTSS all-sky surveys are completed, is needed to test the robustness of these results. In particular, accurate and complete spectroscopic redshift information from WEAVE-LOFAR \citep{2016sf2a.conf..271S} will be especially important to reduce any optical completeness issues in our analysis. Our conclusions also imply that approximately 1 in 40 sources in a flux-density limited survey will be a PS source.

Future very long baseline interferometry (VLBI) observations with the LOFAR international baselines are needed to identify specific high-resolution structures and morphologies for the PS sources in our sample. With this, the environments and symmetries of our sample of PS sources can be investigated to confirm our hypothesis that the majority of these sources are not \textit{frustrated}, but will rather grow to be large radio galaxies. Furthermore, these morphological studies could identify transient, short-lived PS sources by searching for multi-epoch activity.

Detailed studies of the optical hosts of our sample of PS sources are needed to investigate the optical nature of these sources. This can give insight into whether PS sources with quasar and galaxy hosts have different characteristics, which could be suggestive of a different spectral turnover mechanism.

To improve the selection of PS sources in this detection area, a better sensitivity in our low frequency survey is needed, as this is the most limiting factor in identifying PS sources in this study. The final LoLSS data release will enable us to improve this sensitivity, as it will be $\sim$5 times deeper, and have a $\sim$3 times higher resolution than the preliminary data release used in this study (de Gasperin et al., in prep.). The methods of selecting PS sources in LOFAR surveys outlined in this study can then be easily applied to future data releases with these improvements as well as a larger sky coverage of LoTSS and LoLSS to identify a large number of new PS sources. Finally, our empirically derived luminosity functions need to be tested against evolutionary models for radio-loud AGN \citep[][]{2018MNRAS.475.3493B} to ensure those models are consistent with the evolution we are suggesting.

\begin{acknowledgements}

We thank E.\,M.\ Sadler (CASS/University of Sydney) and R.\,D.\,Ekers (CASS) for their helpful suggestions on this work.
JRC thanks the Nederlandse Organisatie voor Wetenschappelijk Onderzoek (NWO) for support via the Talent Programme Veni grant. 
KJD acknowledges funding from the European Union’s Horizon 2020 research and innovation programme under the Marie Sk\l{}odowska-Curie grant agreement No. 892117 (HIZRAD).
MJH acknowledges support from the UK STFC [ST/V000624/1].
LOFAR \citep{2013A&A...556A...2V} is the Low Frequency Array designed and constructed by ASTRON. It has observing, data processing, and data storage facilities in several countries, which are owned by various parties (each with their own funding sources), and that are collectively operated by the ILT foundation under a joint scientific policy. The ILT resources have benefited from the following recent major funding sources: CNRS-INSU, Observatoire de Paris and Université d'Orléans, France; BMBF, MIWF-NRW, MPG, Germany; Science Foundation Ireland (SFI), Department of Business, Enterprise and Innovation (DBEI), Ireland; NWO, The Netherlands; The Science and Technology Facilities Council, UK; Ministry of
Science and Higher Education, Poland; The Istituto Nazionale di Astrofisica (INAF), Italy.
This research made use of the Dutch national e-infrastructure with support of the SURF Cooperative (e-infra 180169) and the LOFAR e-infra group. The Jülich LOFAR Long Term Archive and the German LOFAR network are both coordinated and operated by the Jülich Supercomputing Centre (JSC), and computing resources on the supercomputer JUWELS at JSC were provided by the Gauss Centre for Supercomputing e.V. (grant CHTB00) through the John von Neumann Institute for Computing (NIC).
This research made use of the University of Hertfordshire high-performance computing facility and the LOFAR-UK computing facility located at the University of Hertfordshire and supported by STFC [ST/P000096/1], and of the Italian LOFAR IT computing infrastructure supported and operated by INAF, and by the Physics Department of Turin university (under an agreement with Consorzio Interuniversitario per la Fisica Spaziale) at the C3S Supercomputing Centre, Italy.
This research has made use of the NASA/IPAC Extragalactic Database (NED), which is funded by the National Aeronautics and Space Administration and operated by the California Institute of Technology, the VizieR catalogue access tool, CDS,
Strasbourg, France, and the ``K-corrections calculator'' service available at http://kcor.sai.msu.ru/. This research has also made use of "Aladin sky atlas" developed at CDS, Strasbourg Observatory, France, TOPCAT \citep{2005ASPC..347...29T}, NASA's Astrophysics Data System bibliographic services, and \texttt{Astropy}, a community-developed core Python package for Astronomy \citep{astropy:2018}. This work has also made use of the Python packages \texttt{Numpy} \citep{harris2020array}, \texttt{SciPy} \citep{2020SciPy-NMeth}, and \texttt{Matplotlib} \citep{Hunter:2007}.
\end{acknowledgements}

%
%

\bibliographystyle{aa.bst}
\bibliography{mybib_peaked_spectrum.bib}

\begin{appendix}
\onecolumn
\section{List of column headings for table of PS sample}
The column numbers, names, and units for the columns of the table presenting our PS sample are outlined below.
\label{ap:PS_table}
\begin{table}[h]
    \centering
    \begin{tabularx}{\hsize}{p{0.08\hsize} p{0.12\hsize} p{0.08\hsize} X}
    \hline
    \hline
    Number & Name & Unit & Description \\
    \hline
    1 & LoTSS name & -- & Name of source in the LoTSS catalogue \\
    2 & LoTSS R.A. & degrees & R.A. of the source in the LoTSS catalogue \\
    3 & LoTSS Decl. & degrees & Decl. of the source in the LoTSS catalogue \\
    4 & $S_{\mathrm{LoTSS}}$ & Jy & Integrated LoTSS flux density \\
    5 & $\Delta S_{\mathrm{LoTSs}}$ & Jy & Uncertainty in integrated LoTSS flux density \\
    6 & $a_{\mathrm{low}}$ & Jy & Low frequency amplitude of the power-law fit between 54 and 144\,MHz \\
    7 & \alow & -- & Low frequency spectral index between 54 and 144\,MHz \\
    8 & $\Delta$ \alow & -- & Uncertainty on low frequency spectral index between 54 and 144\,MHz \\
    9 & $a_{\mathrm{high}}$ & Jy & Low frequency amplitude of the power-law fit between 144 and 1400\,MHz \\
    10 & \ahigh & -- & High frequency spectral index between 144 and 1400\,MHz \\
    11 & $\Delta$ \ahigh & -- & Uncertainty on high frequency spectral index between 144 and 1400\,MHz \\
    12 & LoLSS R.A. & degrees & R.A. of the source in the LoLSS catalogue \\
    13 & LoLSS Decl. & degrees & Decl. of the source in the LoLSS catalogue \\
    14 & $S_{\mathrm{LoLSS}}$ & Jy & Integrated LoLSS flux density \\
    15 & $\Delta S_{\mathrm{LoLSS}}$ & Jy & Uncertainty in integrated LoLSS flux density \\
    16 & NVSS R.A. & degrees & R.A. of the source in the NVSS catalogue \\
    17 & NVSS Decl. & degrees & Decl. of the source in the NVSS catalogue \\
    18 & $S_{\mathrm{NVSS}}$ & Jy & Integrated NVSS flux density \\
    19 & $\Delta S_{\mathrm{NVSS}}$ & Jy & Uncertainty in integrated NVSS flux density \\
    20 & TGSS R.A. & degrees & R.A. of the source in the TGSS catalogue \\
    21 & TGSS Decl. & degrees & Decl. of the source in the TGSS catalogue \\
    22 & $S_{\mathrm{TGSS}}$ & Jy & Integrated TGSS flux density \\
    23 & $\Delta S_{\mathrm{TGSS}}$ & Jy & Uncertainty in integrated TGSS flux density \\
    24 & VLSSr R.A. & degrees & R.A. of the source in the VLSSr catalogue \\
    25 & VLSSr Decl. & degrees & Decl. of the source in the VLSSr catalogue \\
    26 & $S_{\mathrm{VLSSr}}$ & Jy & Integrated VLSSr flux density \\
    27 & $\Delta S_{\mathrm{VLSSr}}$ & Jy & Uncertainty in integrated VLSSr flux density \\
    28 & FIRST R.A. & degrees & R.A. of the source in the FIRST catalogue \\
    29 & FIRST Decl. & degrees & Decl. of the source in the FIRST catalogue \\
    30 & $S_{\mathrm{FIRST}}$ & Jy & Integrated FIRST flux density \\
    31 & $\Delta S_{\mathrm{FIRST}}$ & Jy & Uncertainty in integrated FIRST flux density \\
    32 & Inband R.A. & degrees & R.A. of the source in the LoTSS in-band spectrum catalogue \\
    33 & Inband Decl. & degrees & Decl. of the source in the LoTSS in-band spectrum catalogue \\
    34 & $S_{\mathrm{inband\ low}}$ & Jy & Integrated LoTSS 128\,MHz in-band flux density \\
    35 & $\Delta S_{\mathrm{inband\ low}}$ & Jy & Uncertainty in integrated LoTSS 128\,MHz in-band flux density \\
    36 & $S_{\mathrm{inband\ mid}}$ & Jy & Integrated LoTSS 144\,MHz in-band flux density \\
    37 & $\Delta S_{\mathrm{inband\ mid}}$ & Jy & Uncertainty in integrated LoTSS 144\,MHz in-band flux density \\
    38 & $S_{\mathrm{inband\ high}}$ & Jy & Integrated LoTSS 160\,MHz in-band flux density \\
    \hline\hline
    \end{tabularx}
    \label{tab:PS_table}
\end{table}

\begin{table}[]
    \begin{tabularx}{\hsize}{p{0.08\hsize} p{0.12\hsize} p{0.08\hsize} X}
    \multicolumn{4}{l}{continued.}\\
    \hline
    \hline
    Number & Name & Unit & Description \\
    \hline
    39 & $\Delta S_{\mathrm{inband\ high}}$ & Jy & Uncertainty in integrated LoTSS 160\,MHz in-band flux density \\
    40 & Opt. name & -- & Name of source in the optical catalogue \\
    41 & Opt. R.A. & degrees & R.A. of the source in the optical catalogue \\
    42 & Opt. Decl. & degrees & Decl. of the source in the optical catalogue \\
    43 & $z_{\mathrm{spec}}$ & -- & Spectroscopic redshift \\
    44 & $z_{\mathrm{phot}}$ & -- & Photometric redshift \\
    45 & $z_{\mathrm{best}}$ & -- & Best redshift \\
    46 & Ref opt. & -- & Reference of optical catalogue (LoTSS DR1/DR2) \\
    47 & $L_{144\mathrm{MHz}}$ & W\,Hz$^{-1}$ & Calculated 144\,MHz luminosity \\
    48 & $L_{5\mathrm{GHz}}$ & W\,Hz$^{-1}$ & Calculated 5\,GHz luminosity \\
    \hline\hline
    \end{tabularx}
\end{table}

\onecolumn
\section{Peaked-spectrum radio luminosity function - Table}
\label{ap:LF}
\begin{table*}[h]
    \small
    \centering
    \caption{The 144\,MHz luminosity functions for our {SDSS selected} Master sample (left column) and PS sample (right column) in different redshift ranges.}
    \begin{tabular}{c c c | c c c c}
    \hline\hline
    $\log L$ [W Hz$^{-1}$] & $N$ & $\log \Phi$ [Mpc$^{-3}$\,dex$^{-1}$] & $\log L$ [W Hz$^{-1}$] & $N$ & $\log \Phi$ [Mpc$^{-3}$\,dex$^{-1}$] & $\log(\Delta \Phi)$\\
    \hline\hline
    
    \multicolumn{3}{c|}{$z <0.1$; Master Sample} & \multicolumn{4}{c}{$z <0.1$; PS Sample} \\ 
    23.40 $\pm$ 0.4 & 34 & $ -4.71 ^{+ 0.09 }_{- 0.12 }$ & 23.71 $\pm$ 0.5 & 6 & $ -5.31 ^{+ 0.20 }_{- 0.38 }$ & $ 0.31 ^{+ 0.22 }_{- 0.40 }$ \\ 
    24.33 $\pm$ 0.5 & 13 & $ -5.58 ^{+ 0.11 }_{- 0.15 }$ & & & \\
    
    \hline
    \multicolumn{3}{c|}{$0.1 < z <0.5$; Master Sample} & \multicolumn{4}{c}{$0.1 < z <0.5$; PS Sample} \\ 
    23.98 $\pm$ 0.2 & 15 & $ -5.64 ^{+ 0.11 }_{- 0.15 }$    & 25.16 $\pm$ 0.7 & 15 & $ -7.01 ^{+ 0.16 }_{- 0.26 }$& $ 0.92 ^{+ 0.17 }_{- 0.26 }$\\ 
    24.38 $\pm$ 0.2 & 36 & $ -5.90 ^{+ 0.08 }_{- 0.09 }$ & 26.56 $\pm$ 0.7 & 7 & $ -8.01 ^{+ 0.14 }_{- 0.21 }$ & $ 0.85 ^{+ 0.16 }_{- 0.23 }$\\ 
    24.78 $\pm$ 0.2 & 81 & $ -6.09 ^{+ 0.05 }_{- 0.06 }$    & & & \\
    25.18 $\pm$ 0.2 & 151 & $ -6.09 ^{+ 0.04 }_{- 0.04 }$   & & & \\
    25.58 $\pm$ 0.2 & 81 & $ -6.4 ^{+ 0.05 }_{- 0.05 }$ & & &\\ 
    25.98 $\pm$ 0.2 & 34 & $ -6.78 ^{+ 0.07 }_{- 0.08 }$    & & &\\ 
    26.38 $\pm$ 0.2 & 22 & $ -6.97 ^{+ 0.08 }_{- 0.10 }$ & & &\\ 
    26.99 $\pm$ 0.4 & 10 & $ -7.62 ^{+ 0.12 }_{- 0.17 }$    & & &\\
    
    \hline
    \multicolumn{3}{c|}{$0.5 < z <1.0$; Master Sample} & \multicolumn{4}{c}{$0.5 < z <1.0$; PS Sample} \\ 
    25.34 $\pm$ 0.2 & 108 & $ -6.34 ^{+ 0.05 }_{- 0.05 }$ & 25.94 $\pm$ 0.3 & 19 & $ -7.24 ^{+ 0.14 }_{- 0.21 }$ & $ 0.89 ^{+ 0.15 }_{- 0.22 }$ \\ 
    25.74 $\pm$ 0.2 & 305 & $ -6.32 ^{+ 0.04 }_{- 0.04 }$ & 26.54 $\pm$ 0.3 & 13 & $ -7.92 ^{+ 0.12 }_{- 0.17 }$ & $ 1.17 ^{+ 0.13 }_{- 0.17 }$\\ 
    26.14 $\pm$ 0.2 & 314 & $ -6.39 ^{+ 0.03 }_{- 0.03 }$ & 27.14 $\pm$ 0.3 & 8 & $ -8.20 ^{+ 0.13 }_{- 0.19 }$   & $ 0.93 ^{+ 0.16 }_{- 0.22 }$\\ 
    26.54 $\pm$ 0.2 & 133 & $ -6.75 ^{+ 0.04 }_{- 0.04 }$ & 27.74 $\pm$ 0.3 & 3 & $ -8.38 ^{+ 0.24 }_{- 0.42 }$  & $ 0.66 ^{+ 0.25 }_{- 0.44 }$\\ 
    26.94 $\pm$ 0.2 & 61 & $ -7.00 ^{+ 0.07 }_{- 0.08 }$& & &\\ 
    27.34 $\pm$ 0.2 & 21 & $ -7.54 ^{+ 0.09 }_{- 0.11 }$& & &\\ 
    27.92 $\pm$ 0.4 & 18 & $ -7.80 ^{+ 0.10 }_{- 0.13 }$& & & \\
    
    \hline
    \multicolumn{3}{c|}{$1.0 < z <1.5$; Master Sample} & \multicolumn{4}{c}{$1.0 < z <1.5$; PS Sample} \\ 
    25.99 $\pm$ 0.2 & 51 & $ -6.5 ^{+ 0.12 }_{- 0.16 }$ &26.81 $\pm$ 0.4 & 13 & $ -7.7 ^{+ 0.18 }_{- 0.31 }$ & $ 1.45 ^{+ 0.28 }_{- 0.55 }$\\ 
    26.49 $\pm$ 0.2 & 80 & $ -6.45 ^{+ 0.10 }_{- 0.12 }$ & 27.61 $\pm$ 0.4 & 3 & $ -7.97 ^{+ 0.24 }_{- 0.42 }$ & $ 0.95 ^{+ 0.26 }_{- 0.45 }$\\ 
    26.99 $\pm$ 0.2 & 55 & $ -6.15 ^{+ 0.22 }_{- 0.46 }$ & & & \\
    27.49 $\pm$ 0.2 & 38 & $ -6.82 ^{+ 0.12 }_{- 0.16 }$ & & & \\
    28.07 $\pm$ 0.3 & 11 & $ -7.81 ^{+ 0.15 }_{- 0.22 }$ & & & \\
    
    \hline
    \multicolumn{3}{c|}{$1.5 < z <3.0$; Master Sample} & \multicolumn{4}{c}{$1.5 < z <3.0$; PS Sample} \\ 
    26.51 $\pm$ 0.3 & 47 & $ -7.10 ^{+ 0.10 }_{- 0.13 }$& 27.1 $\pm$ 0.5 & 10 & $ -7.89 ^{+ 0.15 }_{- 0.24 }$ & $ 0.90 ^{+ 0.18 }_{- 0.26 }$\\ 
    27.11 $\pm$ 0.3 & 78 & $ -6.98 ^{+ 0.09 }_{- 0.11 }$& 28.1 $\pm$ 0.5 & 6 & $ -9.04 ^{+ 0.16 }_{- 0.25 }$ & $ 1.29 ^{+ 0.20 }_{- 0.31 }$\\ 
    27.71 $\pm$ 0.3 & 49 & $ -7.56 ^{+ 0.09 }_{- 0.12 }$& & &  \\
    28.44 $\pm$ 0.4 & 33 & $ -7.90 ^{+ 0.13 }_{- 0.18 }$& & & \\

    \hline\hline
    \end{tabular}
    \tablefoot{$\log L$ is the centre of each luminosity bin. The number of sources in each respective bin are given by $N$. $\log \Phi$ is the $\log$ luminosity function for each luminosity bin. The errors on the luminosity functions are the Poissonian counting errors. $\log(\Delta\Phi)$ is the offset between the Master sample luminosity function and the PS sample luminosity function at each luminosity bin.}
    \label{tab:LF_z}
\end{table*}

\begin{table*}[h]
    \small
    \centering
    \caption{{The 144\,MHz luminosity functions for our photo-z selected Master sample (left column) and PS sample (right column) in different redshift ranges.}}
    \begin{tabular}{c c c | c c c c}
    \hline\hline
    $\log L$ [W Hz$^{-1}$] & $N$ & $\log \Phi$ [Mpc$^{-3}$\,dex$^{-1}$] & $\log L$ [W Hz$^{-1}$] & $N$ & $\log \Phi$ [Mpc$^{-3}$\,dex$^{-1}$] & $\log(\Delta \Phi)$\\
    \hline\hline
    
    \multicolumn{3}{c|}{$z <0.1$; Master Sample} & \multicolumn{4}{c}{$z <0.1$; PS Sample} \\ 
    23.4 $\pm$ 0.4 & 30 & $ -4.92^{+0.08}_{-0.10}$ & 23.71 $\pm$ 0.5 & 4 & $ -5.61 ^{+ 0.21 }_{- 0.34 }$ & $ 0.48 ^{+ 0.22 }_{- 0.36 }$ \\  
    24.33 $\pm$ 0.5 & 16 & $ -5.57^{+0.10}_{-0.12}$ & & & \\
    
    \hline
    \multicolumn{3}{c|}{$0.1 < z <0.5$; Master Sample} & \multicolumn{4}{c}{$0.1 < z <0.5$; PS Sample} \\ 
    23.98 $\pm$ 0.2 & 30 & $ -5.36^{+0.08}_{-0.10}$  & 25.16 $\pm$ 0.7 & 15 & $ -7.19 ^{+ 0.11 }_{- 0.15 }$ & $ 1.11 ^{+ 0.12 }_{- 0.16 }$ \\  
    24.38 $\pm$ 0.2 & 34 & $ -5.83^{+0.14}_{-0.20}$  & 26.56 $\pm$ 0.7 & 7 & $ -8.01 ^{+ 0.14 }_{- 0.21 }$ & $ 0.87 ^{+ 0.16 }_{- 0.23 }$ \\  
    24.78 $\pm$ 0.2 & 89 & $ -6.05^{+0.05}_{-0.05}$  & & & \\
    25.18 $\pm$ 0.2 & 156 & $ -6.07^{+0.04}_{-0.04}$ & & & \\
    25.58 $\pm$ 0.2 & 96 & $ -6.33^{+0.04}_{-0.05}$  & & &\\ 
    25.98 $\pm$ 0.2 & 29 & $ -6.85^{+0.07}_{-0.09}$  & & &\\ 
    26.38 $\pm$ 0.2 & 27 & $ -6.88^{+0.08}_{-0.09}$  & & &\\ 
    26.99 $\pm$ 0.4 & 7 & $ -7.78^{+0.14}_{-0.21}$   & & &\\
    
    \hline
    \multicolumn{3}{c|}{$0.5 < z <1.0$; Master Sample} & \multicolumn{4}{c}{$0.5 < z <1.0$; PS Sample} \\ 
    25.34 $\pm$ 0.2 & 158 & $ -6.22 ^{+ 0.04 }_{- 0.04 }$ & 25.94 $\pm$ 0.3 & 12 & $ -7.34 ^{+ 0.17 }_{- 0.29 }$ & $ 1.07 ^{+ 0.17 }_{- 0.29 }$ \\  
    25.74 $\pm$ 0.2 & 340 & $ -6.23 ^{+ 0.04 }_{- 0.04 }$ & 26.54 $\pm$ 0.3 & 14 & $ -7.91 ^{+ 0.11 }_{- 0.14 }$ & $ 1.29 ^{+ 0.11 }_{- 0.15 }$ \\ 
    26.14 $\pm$ 0.2 & 324 & $ -6.33 ^{+ 0.03 }_{- 0.03 }$ & 27.14 $\pm$ 0.3 & 6 & $ -8.30 ^{+ 0.15 }_{- 0.23 }$ & $ 1.11 ^{+ 0.18 }_{- 0.26 }$ \\  
    26.54 $\pm$ 0.2 & 157 & $ -6.62 ^{+ 0.04 }_{- 0.04 }$ & 27.74 $\pm$ 0.3 & 1 & $ -9.12 ^{+ 0.36 }_{- 0.99 }$ & $ 1.32 ^{+ 0.38 }_{- 1.01 }$ \\ 
    26.94 $\pm$ 0.2 & 76 & $ -6.92 ^{+ 0.05 }_{- 0.06 }$  & & &\\ 
    27.34 $\pm$ 0.2 & 22 & $ -7.46 ^{+ 0.09 }_{- 0.12 }$  & & &\\ 
    27.92 $\pm$ 0.4 & 13 & $ -7.96 ^{+ 0.11 }_{- 0.16 }$  & & & \\
    
    \hline
    \multicolumn{3}{c|}{$1.0 < z <1.5$; Master Sample} & \multicolumn{4}{c}{$1.0 < z <1.5$; PS Sample} \\ 
    25.99 $\pm$ 0.2 & 54 & $ -5.98 ^{+ 0.21 }_{- 0.40 }$ & 26.81 $\pm$ 0.4 & 10 & $ -8.0 ^{+ 0.13 }_{- 0.20 }$ & $ 1.6 ^{+ 0.17 }_{- 0.23 }$ \\  
    26.49 $\pm$ 0.2 & 99 & $ -5.98 ^{+ 0.14 }_{- 0.20 }$ & 27.61 $\pm$ 0.4 & 3 & $ -7.83 ^{+ 0.24 }_{- 0.42 }$ & $ 0.87 ^{+ 0.31 }_{- 0.57 }$ \\  
    26.99 $\pm$ 0.2 & 61 & $ -6.64 ^{+ 0.10 }_{- 0.13 }$ & & & \\
    27.49 $\pm$ 0.2 & 24 & $ -6.75 ^{+ 0.20 }_{- 0.38 }$ & & & \\
    28.07 $\pm$ 0.3 & 13 & $ -7.79 ^{+ 0.13 }_{- 0.18 }$ & & & \\
    
    \hline
    \multicolumn{3}{c|}{$1.5 < z <3.0$; Master Sample} & \multicolumn{4}{c}{$1.5 < z <3.0$; PS Sample} \\ 
    26.51 $\pm$ 0.3 & 56 & $ -7.25 ^{+ 0.10 }_{- 0.14 }$ & 27.1 $\pm$ 0.5 & 7 & $ -8.34 ^{+ 0.22 }_{- 0.45 }$ & $ 1.25 ^{+ 0.24 }_{- 0.47 }$ \\  
    27.11 $\pm$ 0.3 & 94 & $ -7.08 ^{+ 0.09 }_{- 0.12 }$& 28.1 $\pm$ 0.5 & 3 & $ -9.16 ^{+ 0.24 }_{- 0.42 }$ & $ 1.38 ^{+ 0.25 }_{- 0.43 }$ \\
    27.71 $\pm$ 0.3 & 65 & $ -7.38 ^{+ 0.16 }_{- 0.26 }$& & &  \\
    28.44 $\pm$ 0.4 & 34 & $ -8.13 ^{+ 0.08 }_{- 0.09 }$& & & \\

    \hline\hline
    \end{tabular}
    \tablefoot{$\log L$ is the centre of each luminosity bin. The number of sources in each respective bin are given by $N$. $\log \Phi$ is the $\log$ luminosity function for each luminosity bin. The errors on the luminosity functions are the Poissonian counting errors. $\log(\Delta\Phi)$ is the offset between the Master sample luminosity function and the PS sample luminosity function at each luminosity bin.}
    \label{tab:LF_z2}
\end{table*}

\end{appendix}

\end{document}